\documentclass[%
 reprint,
 amsmath,amssymb,
 aps,
]{revtex4-2}

\usepackage{graphicx}
\usepackage{dcolumn}
\usepackage{bm}
\usepackage{hyperref}

\usepackage{slashed}
\usepackage{float}
\usepackage{xcolor}

\begin{document}


\title{Tree-Level Interaction Vertices in the Linear Sigma Model with Quarks under Global Rotation}%

\author{Luis A. Hernández}
 \affiliation{Departamento de F\'isica, Universidad Aut\'onoma Metropolitana-Iztapalapa, Avenida San Rafael Atlixco 186, Ciudad de México 09340, Mexico.}
 
\author{Eduardo Lazcano}%
 \affiliation{Departamento de F\'isica, Universidad Aut\'onoma Metropolitana-Iztapalapa, Avenida San Rafael Atlixco 186, Ciudad de México 09340, Mexico.}

\author{R. Zamora}
\affiliation{Centro de Investigaci\'on y Desarrollo en Ciencias Aeroespaciales (CIDCA), Academia Politécnica Aeronáutica, Fuerza A\'erea de Chile, Casilla 8020744, Santiago, Chile}
\affiliation{Facultad de Ingenier\'ia y Arquitectura, Universidad Central de Chile, Av. Sta. Isabel 1186, Santiago 8330601, Chile.}



\date{\today}

\begin{abstract}
We derive the tree-level interaction vertices of the linear sigma model with quarks in the presence of global rotation. Starting from the scalar and fermionic field solutions in a rotating background, we construct the corresponding spectral representations and use them to evaluate the mesonic self-interactions and Yukawa couplings of the model. We show that rotation modifies the usual momentum-space structure of the interaction vertices: conservation of longitudinal momentum is accompanied by angular-momentum selection rules, while the transverse dynamics is encoded in rotation-induced radial form factors involving products of Bessel functions. The resulting expressions provide a set of interaction rules adapted to the symmetries and finite geometry of the rotating system. Although derived within the linear sigma model with quarks, the structures obtained for scalar self-interactions and scalar-fermion couplings can be extended to a broader class of quantum field theories under global rotation. These results provide the building blocks for perturbative calculations and contribute toward a microscopic quantum-field-theoretical description of interacting matter in rotating environments.
\end{abstract}

\maketitle


\section{\label{sec1} Introduction}
In relativistic heavy-ion collisions, particularly in non-central events, extremely large rotational motion has been reported~\cite{Deng:2016gyh,Deng:2020ygd}, with a global vorticity corresponding to an angular velocity of about $\Omega \simeq (9 \pm 1) \times 10^{21} \ \text{s}^{-1}$\cite{STAR:2017ckg}.
This rotational dynamics enriches the description of the reaction processes generated in heavy-ion collisions and has become a topic of considerable interest within the high-energy physics community~\cite{Huang:2020dtn,Tsegelnik:2022eoz,Gao:2023wwo,Huang:2024aob,Ayala:2025kuc,Savchuk:2025kuk,Sahoo:2026lrw,Padhan:2026mwg}. One of the most extensively studied experimental observables associated with global vorticity is hadron polarization~\cite{Xie:2015xpa,Xie:2016fjj,Becattini:2016gvu,Sorin:2016smp,Li:2017slc,Sun:2017xhx,Han:2017hdi,Xia:2018tes,Baznat:2017jfj,Karpenko:2018erl,Suvarieva:2018wmh,Kolomeitsev:2018svb,Xie:2019jun,Ayala:2020soy,Ayala:2021xrn,Ayala:2023xyn,STAR:2023nvo,Palermo:2024tza,Sapna:2025yss,Ivanov:2025izv,STAR:2025dgs,Troshin:2025fko}. On the theoretical side, considerable efforts have been devoted to understanding the impact of rotation on the thermodynamic properties of the system, particularly regarding the QCD phase transition and how rotation modifies the determination of the transition line~\cite{Jiang:2016wvv,Chernodub:2016kxh,Wang:2018sur,Braguta:2020biu,Chen:2020ath,Chernodub:2020qah,Braguta:2021jgn,Fujimoto:2021xix,Chernodub:2022veq,Braga:2022yfe,Gaspar:2023nqk,Chen:2023cjt,Yang:2023vsw,TabatabaeeMehr:2023tpt,Sun:2023kuu,Cao:2023olg,Jiang:2023zzu,Braguta:2023iyx,Sun:2024anu,Wang:2024szr,Hernandez:2024nev,Siri:2025kdw,Farias:2025vss,Zhu:2025pxh,Hernandez:2026lyd}.
Other relevant developments include the implementation of vorticity in hydrodynamic models~\cite{Florkowski:2018fap,Sahoo:2023xnu,Palermo:2023cup,Xie:2023gbo,Florkowski:2024cif,Huang:2024ffg,Bhadury:2024ckc,Fang:2025aig}, to describe the thermally equilibrated stages of the reaction, as well as microscopic studies aimed at understanding the propagation of fields in rotating media and the relaxation times associated with spin alignment of quarks and gluons in a thermal environment~\cite{Ayala:2019iin,Kapusta:2019sad,Ayala:2020ndx,Ayala:2023vgv}.

In order to achieve a microscopic description of the dynamics generated in non-central relativistic heavy-ion collisions, it is essential to understand how the rotational motion of the medium affects the fundamental interactions among its constituents. Within the framework of Quantum Field Theory, these interactions are encoded in the structure of the interaction vertices, which determine the elementary scattering and decay processes of the theory. In the presence of a globally rotating background, the equations of motion for scalar and fermionic fields are modified by the underlying geometry, leading to solutions that are no longer described by plane waves. As a consequence, the interaction vertices acquire a non-trivial space-time structure and, after transforming to momentum space, the usual point-like energy-momentum conservation expressed through Dirac delta functions is replaced by distributed structures that naturally emerge as rotation-induced form factors. In this work, we derive the tree-level interaction vertices of the linear sigma model with quarks in the presence of global rotation, providing a microscopic framework to study how rotation modifies the elementary interactions within strongly interacting matter.

This work is organized as follows. In Sec.~\ref{sec2}, we introduce the linear sigma model with quarks, discussing its main features, field content, and notation. In Sec.~\ref{sec3}, we present the equations of motion for scalar and fermionic fields in the presence of global rotation and summarize their corresponding solutions. In Sec.~\ref{sec4}, we derive the tree-level interaction vertices in a rotating background and show how rotation-induced form factors naturally emerge in momentum space. Finally, in Sec.~\ref{sec5}, we present our conclusions.

\section{\label{sec2} Linear Sigma Model with quarks}

The linear sigma model with quarks (LSMq) is a renormalizable effective theory that captures some of the low-energy properties of QCD while preserving chiral symmetry and allowing for its spontaneous breaking. The model contains scalar and pseudoscalar mesonic degrees of freedom together with the two lightest quark flavors. The Lagrangian density is given by
\begin{align}
    \mathcal{L} &= \frac{1}{2}(\partial_{\mu}\sigma)^2+\frac{1}{2}(\partial_{\mu} \pi_{0})^2+\partial_{\mu} \pi_- \partial^{\mu} \pi_{+}\nonumber \\
    &+\frac{a^2}{2}(\sigma^2+\pi_0^2+2 \pi_{-}\pi_{+})-\frac{\lambda}{4} \left( \sigma^2+\pi_0^2+2 \pi_{-}\pi_{+}\right)^2 \nonumber \\
    &+i\bar{\psi} \slashed{\partial}\psi - g\bar{\psi} ( \sigma+i \gamma^{5} \vec{\tau}\cdot\vec{\pi} ) \psi,
\label{lagrangian}
\end{align}
where $\psi$ denotes the $SU(2)$ isospin doublet of up and down quarks, $\sigma$ is an isospin singlet scalar field, and $\{\pi_+,\pi_-,\pi_0\}$ represents the pion isotriplet. The Pauli matrices $\tau_i$ act in isospin space, whereas $\lambda$ and $g$ correspond to the boson self-coupling and Yukawa coupling constants, respectively. The parameter $a^2$ is the squared mass parameter.

In the spontaneously broken phase, the $\sigma$ field develops a vacuum expectation value $v$, implemented through the shift
\begin{equation}
\sigma \rightarrow \sigma + v.
\label{shift}
\end{equation}
After performing this shift, the Lagrangian can be separated into free and interacting contributions
\begin{align}
    \mathcal{L}&=\frac{1}{2}\partial_{\mu}\sigma \partial^{\mu}\sigma+\frac{1}{2}\partial_{\mu}\pi_{0}\partial^{\mu}\pi_{0}+\partial_{\mu}\pi_{-}\partial^{\mu}\pi_{+}\nonumber\\
    &-\frac{1}{2}m_{\sigma}^{2}\sigma^{2}-\frac{1}{2}m_{\pi}^{2}\pi_{0}^{2}-m_{\pi}^{2}\pi_{-}\pi_{+}+i\bar{\psi}\slashed{\partial}\psi\nonumber\\
    &-m_{f}\bar{\psi}\psi+\frac{a^2}{2}v^2-\frac{\lambda}{4}v^4-\lambda v^{3}\sigma+a^{2}v\sigma +\mathcal{L}_{int}.
\label{linearsigmamodelSSB}
\end{align}
The tree-level masses are given by
\begin{equation}
    m_{\sigma}^{2}=3\lambda v^2-a^2, \qquad
    m_{\pi}^{2}=\lambda v^2-a^2, \qquad
    m_{f}=gv,
\label{masses}
\end{equation}
where the vacuum expectation value $v$ plays the role of the order parameter associated with spontaneous chiral symmetry breaking.

The interaction sector of the theory is encoded in the interacting Lagrangian density
\begin{align}
    \mathcal{L}_{int}&=-\frac{\lambda}{4}\sigma^{4}-\lambda v\sigma^{3}-\lambda\sigma^{2}\pi_{-}\pi_{+} -2\lambda v \sigma\pi_{-}\pi_{+} \nonumber \\
    &-\frac{\lambda}{2}\sigma^{2}\pi_{0}^{2}-\lambda v\sigma \pi_{0}^{2}-\lambda \pi_{-}^{2}\pi_{+}^{2}-\lambda\pi_{-}\pi_{+}\pi_{0}^{2}-\frac{\lambda}{4}\pi_{0}^{4} \nonumber \\
    &-g\bar{\psi}\psi\sigma-ig\gamma^{5}\bar{\psi}\left(\tau_{+}\pi_{+}+\tau_{-}\pi_{-}+\tau_{3}\pi_{0}\right)\psi,
\label{interactinglagrangian}
\end{align}
from which the interaction vertices of the theory are obtained. In the standard formulation of perturbative Quantum Field Theory, these terms generate local interaction vertices together with the corresponding momentum-conserving Dirac delta functions. As will be shown in the following sections, the presence of a globally rotating background modifies the structure of the field solutions and consequently alters the momentum-space representation of the interaction vertices.


\section{\label{sec3} Field Solutions under Rotation: Scalars and Spinors}

The effects of global rotation are incorporated through a rotating background geometry. The interaction region produced after a non-central relativistic heavy-ion collision is modeled as a rigid cylinder rotating around the $\hat{z}$ axis with constant angular velocity $\Omega$. The corresponding metric tensor is given by
\begin{equation}
    g_{\mu\nu} =
    \begin{pmatrix}
    1-(x^2+y^2)\Omega^2 & y \Omega & -x \Omega & 0 \\
    y \Omega & -1 & 0 & 0 \\
    -x \Omega & 0 & -1 & 0 \\
    0 & 0 & 0 & -1
    \end{pmatrix}.
\label{eq:metric}
\end{equation}
Within this background, fermionic fields are described by the Dirac equation in curved space-time
\begin{equation}
    \left[i(\gamma^\mu \partial_\mu +\Gamma_\mu)-m\right]\psi = 0,
\label{eq:DiracEc1}
\end{equation}
where $\Gamma_\mu$ denotes the spin connection and $\gamma^\mu=e^\mu_i\gamma^i$ are the Dirac matrices in the rotating frame.

Using the tetrad formalism in the Cartesian gauge~\cite{Ambrus:2014uqa}, together with the corresponding Christoffel symbols associated with the metric in Eq.~(\ref{eq:metric}), the spin connection reduces to
\begin{equation}
    \Gamma_\mu \to \Gamma_t=-\frac{i}{2}\Omega \sigma^{12},
\end{equation}
while the gamma matrices in the rotating frame become
\begin{eqnarray}
    \gamma^t &=& \gamma^0,\nonumber \\
    \gamma^x &=& \gamma^1+y\Omega\gamma^0,\nonumber \\
    \gamma^y &=& \gamma^2-x\Omega\gamma^0,\nonumber \\
    \gamma^z &=& \gamma^3.
\end{eqnarray}

Therefore, the Dirac equation can be written as
\begin{equation}
    \left[\gamma^0(i\partial_t+\Omega \hat{J}_z)+i\vec{\gamma}\cdot\vec{\nabla}-m \right]\psi=0,
\label{DiracEc3}
\end{equation}
where
\begin{equation}
    \hat{J}_z=\hat{L}_z+\hat{S}_z.
\end{equation}
$\hat{J}_z$ denotes the $\hat{z}$-component of the total angular momentum operator, where the first term corresponds to the orbital angular momentum ($\hat{L}_z$), while the second corresponds to the spin angular momentum, ($\hat{S}_z$). Equation~(\ref{DiracEc3}) explicitly shows that the rotational background couples to the total angular momentum of the fermionic field. Consequently, the eigenfunctions of the system are no longer plane waves but acquire a non-trivial spatial structure determined by the rotational geometry.

The complete derivation of the fermionic solutions in a rotating background has been presented in previous works~\cite{Ayala:2021osy}. For the purposes of the present analysis, we only summarize the main results relevant for the construction of the interaction vertices.

In order to find the solution of Eq.~(\ref{DiracEc3}), we propose the solution
\begin{equation}
    \psi(x) =  \left[\gamma^0(i\partial_t+\Omega \hat{J}_z)+i\vec{\gamma}\cdot \vec{\nabla}+m \right]\phi(x).
    \label{Ansantz}
\end{equation}
Substituting Eq.~(\ref{Ansantz}) into Eq.~(\ref{DiracEc3}), we obtain the scalar differential equation
\begin{equation}
    \left[(i\partial_t+\Omega \hat{J}_z)^2 + \partial_x^2 + \partial_y^2+\partial_z^2 -m^2 \right]\phi(x)=0,
\end{equation}
which has the structure of a Klein-Gordon-type equation. Since this expression does not contain matrices, we can rewrite it in cylindrical coordinates, and therefore obtain
\begin{align}
    \bigg[(i\partial_t+\Omega \hat{J}_z)^2 &+ \left(\partial_r^2+ \frac{1}{r}\partial_r + \frac{1}{r^2}\partial_\varphi^2\right) \nonumber \\
    & +\partial_z^2-m^2 \big]\phi(x)=0.
\label{KGcylindrical}    
\end{align}
Equation~(\ref{KGcylindrical}) provides the key structure governing the field solutions in the rotating background. For fermions, the total angular momentum operator contains both orbital and spin contributions, namely
\begin{equation}
    \hat{J}_z=-i\partial_\varphi+\hat{S}_z.
\end{equation}
For scalar fields, however, the spin contribution vanishes, $\hat{S}_z=0$, and therefore the corresponding Klein-Gordon equation in the rotating background becomes
\begin{align}
    \bigg[(i\partial_t+\Omega \hat{L}_z)^2
    &+\left(\partial_r^2+\frac{1}{r}\partial_r+\frac{1}{r^2}\partial_\varphi^2\right) \nonumber \\
    &+\partial_z^2-m^2\big]\phi_s(x)=0,
\label{KGscalar}
\end{align}
with $\hat{L}_z=-i\partial_\varphi$.

Assuming separable solutions of the form
\begin{equation}
    \phi_s(x)=e^{-iEt+ik_z z+il\varphi}R(r),
\end{equation}
the radial equation becomes
\begin{equation}
    \left[
    r^2\partial_r^2+r\partial_r+
    (r^2k_\perp^2-l^2)
    \right]R(r)=0,
\end{equation}
where
\begin{equation}
    k_\perp^2=(E+l\Omega)^2-k_z^2-m^2.
\end{equation}
Regularity at the origin selects Bessel functions of the first kind, such that the scalar solutions are given by
\begin{equation}
    \phi_s(x)=
    J_l(k_\perp r)\,
    e^{-iEt+ik_z z+il\varphi}.
\label{ScalarSolution}
\end{equation}

For fermionic fields, the spin contribution in $\hat{J}_z=-i\partial_\varphi+\hat{S}_z$ leads to a richer angular structure. Since the eigenvalues of $\hat S_z$ are $s=\pm 1/2$, the conservation of the total angular momentum $j=l+s$ requires the upper and lower spinor components to carry angular momenta differing by one unit. Thus, writing
\begin{equation}
    \phi(x)=e^{-iEt+ik_z z+il\varphi}
    \begin{pmatrix}
    R_1(r)\\
    e^{i\varphi}R_2(r)\\
    R_3(r)\\
    e^{i\varphi}R_4(r)
    \end{pmatrix},
\end{equation}
Eq.~(\ref{KGcylindrical}) reduces to Bessel equations for the radial functions. Regularity at the origin selects
\begin{equation}
    R_{1,3}(r)=J_l(k_\perp r),
    \qquad
    R_{2,4}(r)=J_{l+1}(k_\perp r),
\end{equation}
with
\begin{equation}
    k_\perp^2=\tilde E^2-k_z^2-m^2,
    \qquad
    \tilde E=E+j\Omega .
\end{equation}
Therefore, the auxiliary spinor $\phi(x)$ can be written as
\begin{equation}
    \phi(x)=
    \begin{pmatrix}
    J_l(k_\perp r)\\
    J_{l+1}(k_\perp r)e^{i\varphi}\\
    J_l(k_\perp r)\\
    J_{l+1}(k_\perp r)e^{i\varphi}
    \end{pmatrix}
    e^{-iEt+ik_z z+il\varphi}.
\label{AuxSpinorSolution}
\end{equation}
Having determined the solutions $\phi$, Eq.~(\ref{Ansantz}) can be used to find the spinor wave functions which become
\begin{eqnarray}\nonumber
\psi(x) &=& \begin{pmatrix}
\tilde{E}+m & 0 & -k_z & -\mathcal{P}_{-} \\
0 & \tilde{E}+m & -\mathcal{P}_{+} & k_z \\
k_z & \mathcal{P}_{-} & -\tilde{E}+m & 0 \\
\mathcal{P}_{+} & -k_z & 0 & -\tilde{E}+m\end{pmatrix} \\ \label{eq:psiphi2}
&\times& 
\begin{pmatrix}
J_l(k_\perp r) \\
J_{l+1}(k_\perp r)e^{i\varphi} \\
J_l(k_\perp r) \\
J_{l+1}(k_\perp r)e^{i\varphi}\end{pmatrix} e^{-iEt+ik_z z +il \varphi }\;,
\end{eqnarray}
where $\mathcal{P}_{\pm}=k_x\pm i k_y$. In cylindrical coordinates
\begin{equation}
    \mathcal{P}_{\pm}=-i e^{\pm i\varphi}
    \left(\partial_r \pm\frac{i}{r}\partial_\varphi\right),
\end{equation}
which satisfy
\begin{equation}
    \mathcal{P}_{\pm}
    \left[e^{il\varphi}J_l(k_\perp r)\right]=\pm i k_\perp
    e^{i(l\pm1)\varphi}J_{l\pm1}(k_\perp r),
\end{equation}
we obtain the explicit spinor solution
\begin{align}
    \psi(x)&=
    \begin{pmatrix}
    [\tilde E+m-k_z+i k_\perp]J_l(k_\perp r)\\
    [\tilde E+m+k_z-i k_\perp]J_{l+1}(k_\perp r)e^{i\varphi}\\
    [-\tilde E+m+k_z-i k_\perp]J_l(k_\perp r)\\
    [-\tilde E+m-k_z+i k_\perp]J_{l+1}(k_\perp r)e^{i\varphi}
    \end{pmatrix} \nonumber \\
    & \times e^{-iEt+ik_z z+il\varphi}.
\label{SpinorSolution}
\end{align}

These scalar and fermionic solutions provide the external field modes required to construct the interaction vertices in coordinate space. As shown in the next section, their non-plane-wave structure is responsible for the emergence of rotation-induced form factors in momentum space.


\section{\label{sec4} Modified Feynman Rules in a Rotating Background}

We now derive the interaction vertices of the LSMq in the rotating background. For clarity, we organize them into three classes: four-meson vertices, three-meson vertices, and Yukawa-type vertices involving mesons and quarks.

\begin{figure}[b]
    \centering
    \includegraphics[scale=0.65]{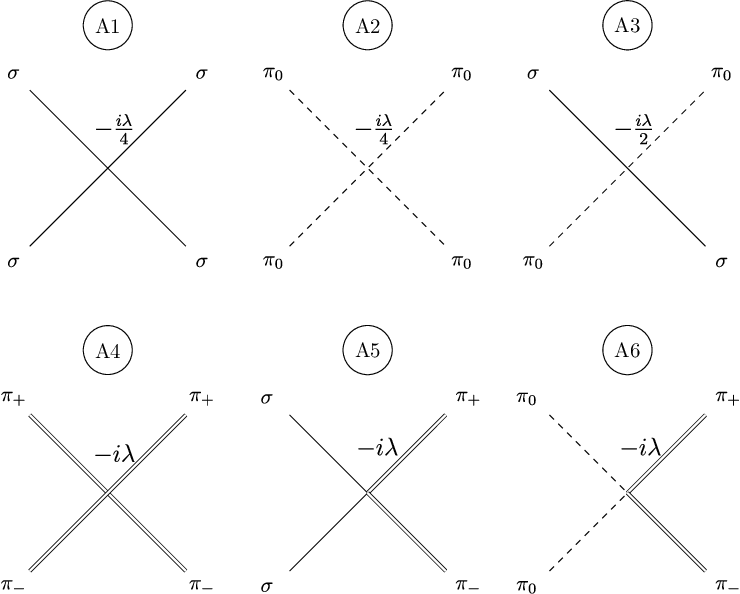}
    \caption{Feynman diagrams corresponding to the vertices with four meson fields interacting. Solid, dashed and double lines denote the sigma meson, the neutral pion and the charged pions, respectively.}
    \label{fig1}
\end{figure}

The Feynman diagrams associated with the four-meson interaction vertices are shown in Fig.~\ref{fig1}. They comprise six independent structures, labeled $Ai$, with $i=1,\ldots,6$. We begin with $A1$ which describes the self-interaction of four sigma fields. From the corresponding interaction term in the Lagrangian density, Eq.~(\ref{interactinglagrangian}), the vertex function is given by
\begin{equation}
    V_{A1}=\langle\sigma_{p_{4}}\sigma_{p_{3}}|i\int d^{4}x\left(-\frac{\lambda\sigma^{4}(x)}{4}\right)|\sigma_{p_{2}}\sigma_{p_{1}}\rangle,
    \label{vertexsigmainitial}
\end{equation}
where the one-particle states and the sigma field are expressed as
\begin{equation}
    |\sigma\rangle=\sqrt{2E_{p}^{(\sigma)}}a^{\dagger}_{\vec{p}}|0\rangle,
    \label{ketsfromvacuum}
\end{equation}
and
\begin{equation}
    \sigma=\int\frac{d^{3}p}{\left(2\pi\right)^{3}}\frac{1}{\sqrt{2E_{p}^{(\sigma)}}}\left(a_{\vec{p}}\,f\left(\vec{x}, \vec{p}\right)+a_{\vec{p}}^{\dagger}\,g\left(\vec{x},\vec{p}\right)\right),
    \label{spectralrepsigma}
\end{equation}
respectively, with
\begin{equation}
    E_{p}^{(\sigma)}=\sqrt{p^{2}_{\perp}+p_{z}^{2}+m_\sigma^2}-\Omega l_{p},
    \label{EnergySigmaMeson}
\end{equation}
and
\begin{align}
    f(\vec{x},\vec{p})&= J_{l_p}\left(rp_{\perp}\right)e^{i\left(-E_{p}t+p_z z+l_p\varphi\right)}, \nonumber \\
    g(\vec{x},\vec{p})&= J_{l_p}\left(rp_{\perp}\right)e^{i\left(E_{p}t-p_z z-l_p\varphi\right)}.
    \label{inoutwaves}
\end{align}
Here, $a_{\vec{p}}^{\dagger}$ and $a_{\vec{p}}$ are the creation and annihilation operators, respectively. The functions $f(\vec{x},\vec{p})$ and $g(\vec{x},\vec{p})$ correspond to the incoming and outgoing scalar modes obtained from Eq.~(\ref{ScalarSolution}). In cylindrical coordinates $(r,\varphi,z)$, the integer $l_p$ labels the eigenvalue of the angular momentum along the rotation axis.

Substituting Eqs.~(\ref{ketsfromvacuum}) and~(\ref{spectralrepsigma}) into Eq.~(\ref{vertexsigmainitial}), we obtain
\begin{align}
    V_{A1}&=-\frac{i\lambda}{4}\int\frac{d^{4}x\,d^{3}q_{1}\,d^{3}q_{2}\,d^{3}q_{3}\,d^{3}q_{4}}{(2\pi)^{12}4\sqrt{E_{q_1}^{(\sigma)}E_{q_2}^{(\sigma)}E_{q_3}^{(\sigma)}E_{q_4}^{(\sigma)}}}\nonumber \\
    &\times 4\sqrt{E_{p_1}^{(\sigma)}E_{p_2}^{(\sigma)}E_{p_3}^{(\sigma)}E_{p_4}^{(\sigma)}}\nonumber \\
&\times \langle0|N\bigg[a_{\vec{p}_{4}}a_{\vec{p}_{3}}\Big(a_{\vec{q}_{1}}f_{1}+a_{\vec{q}_{1}}^{\dagger}g_{1}\Big) \Big(a_{\vec{q}_{2}}f_{2}+a_{\vec{q}_{2}}^{\dagger}g_{2}\Big) \nonumber \\
&\times \Big(a_{\vec{q}_{3}}f_{3}+a_{\vec{q}_{3}}^{\dagger}g_{3}\Big)\Big(a_{\vec{q}_{4}}f_{4}+a_{\vec{q}_{4}}^{\dagger}g_{4}\Big)a_{\vec{p}_{2}}^{\dagger}a_{\vec{p}_{1}}^{\dagger}\bigg]|0\rangle,
\end{align}
where $N$ denotes normal ordering. For compactness, we have introduced the notation $f_{i}=f\left(\vec{x}, \vec{q}_{i}\right)$ and $g_{i}=g\left(\vec{x},\vec{q}_{i}\right)$. Expanding the product and retaining only the terms that contribute to the matrix element, we find
\begin{align}
        V_{A1}&=-\frac{i\lambda}{4}\int\frac{d^{4}x\,d^{3}q_{1}\,d^{3}q_{2}\,d^{3}q_{3}\,d^{3}q_{4}}{(2\pi)^{12}}\nonumber \\ &\times \sqrt{\frac{E_{p_1}^{(\sigma)}E_{p_2}^{(\sigma)}E_{p_3}^{(\sigma)}E_{p_4}^{(\sigma)}}{E_{q_1}^{(\sigma)}E_{q_2}^{(\sigma)}E_{q_3}^{(\sigma)}E_{q_4}^{(\sigma)}}}\nonumber \\
        &\times \langle0|\bigg[a_{\vec{p}_{4}}a_{\vec{p}_{3}}a_{\vec{q}_{1}}a_{\vec{q}_{2}}a_{\vec{q}_{3}}^{\dagger}a_{\vec{q}_{4}}^{\dagger}a_{\vec{p}_{2}}^{\dagger}a_{\vec{p}_{1}}^{\dagger}f_{1}f_{2}g_{3}g_{4}\nonumber \\ &+a_{\vec{p}_{4}}a_{\vec{p}_{3}}a_{\vec{q}_{1}}a_{\vec{q}_{2}}^{\dagger}a_{\vec{q}_{3}}a_{\vec{q}_{4}}^{\dagger}a_{\vec{p}_{2}}^{\dagger}a_{\vec{p}_{1}}^{\dagger}f_{1}g_{2}f_{3}g_{4}\nonumber \\ &+a_{\vec{p}_{4}}a_{\vec{p}_{3}}a_{\vec{q}_{1}}a_{\vec{q}_{2}}^{\dagger}a_{\vec{q}_{3}}^{\dagger}a_{\vec{q}_{4}}a_{\vec{p}_{2}}^{\dagger}a_{\vec{p}_{1}}^{\dagger}f_{1}g_{2}g_{3}f_{4}\nonumber \\ &+a_{\vec{p}_{4}}a_{\vec{p}_{3}}a_{\vec{q}_{1}}^{\dagger}a_{\vec{q}_{2}}a_{\vec{q}_{3}}a_{\vec{q}_{4}}^{\dagger}a_{\vec{p}_{2}}^{\dagger}a_{\vec{p}_{1}}^{\dagger}g_{1}f_{2}f_{3}g_{4}\nonumber \\ &+a_{\vec{p}_{4}}a_{\vec{p}_{3}}a_{\vec{q}_{1}}^{\dagger}a_{\vec{q}_{2}}a_{\vec{q}_{3}}^{\dagger}a_{\vec{q}_{4}}a_{\vec{p}_{2}}^{\dagger}a_{\vec{p}_{1}}^{\dagger}g_{1}f_{2}g_{3}f_{4}\nonumber \\ &+a_{\vec{p}_{4}}a_{\vec{p}_{3}}a_{\vec{q}_{1}}^{\dagger}a_{\vec{q}_{2}}^{\dagger}a_{\vec{q}_{3}}a_{\vec{q}_{4}}a_{\vec{p}_{2}}^{\dagger}a_{\vec{p}_{1}}^{\dagger}g_{1}g_{2}f_{3}f_{4}\Bigg]|0\rangle.
\end{align}
Using the canonical commutation relations and normal ordering the operators, we obtain
\begin{widetext}
\begin{align}
        V_{A1}&=-\frac{i\lambda}{4}\int\frac{d^{4}x\,d^{3}q_{1}\,d^{3}q_{2}\,d^{3}q_{3}\,d^{3}q_{4}}{(2\pi)^{12}}\sqrt{\frac{E_{p_1}^{(\sigma)}E_{p_2}^{(\sigma)}E_{p_3}^{(\sigma)}E_{p_4}^{(\sigma)}}{E_{q_1}^{(\sigma)}E_{q_2}^{(\sigma)}E_{q_3}^{(\sigma)}E_{q_4}^{(\sigma)}}} (2\pi)^{12}\nonumber \\ 
        &\times \left\{ \left( \delta^{(3)}(\vec{p}_3-\vec{q}_1)\delta^{(3)}(\vec{p}_4-\vec{q}_2)+\delta^{(3)}(\vec{p}_3-\vec{q}_2)\delta^{(3)}(\vec{p}_4-\vec{q}_1) \right)\left( \delta^{(3)}(\vec{p}_1-\vec{q}_3)\delta^{(3)}(\vec{p}_2-\vec{q}_4)+\delta^{(3)}(\vec{p}_2-\vec{q}_3)\delta^{(3)}(\vec{p}_1-\vec{q}_4) \right)\right. \nonumber \\
    &\left. \ \ \ \ \ \times g_1g_2f_3f_4 \right. \nonumber \\
    &\left. +\left(\delta^{(3)}(\vec{p}_3-\vec{q}_1)\delta^{(3)}(\vec{p}_4-\vec{q}_3)+\delta^{(3)}(\vec{p}_3-\vec{q}_3)\delta^{(3)}(\vec{p}_4-\vec{q}_1)\right)\left( \delta^{(3)}(\vec{p}_1-\vec{q}_2)\delta^{(3)}(\vec{p}_2-\vec{q}_4)+\delta^{(3)}(\vec{p}_2-\vec{q}_2)\delta^{(3)}(\vec{p}_1-\vec{q}_4) \right)\right. \nonumber \\
    &\left. \ \ \ \ \ \times g_1f_2g_3f_4 \right. \nonumber \\
        &\left. +\left( \delta^{(3)}(\vec{p}_3-\vec{q}_1)\delta^{(3)}(\vec{p}_4-\vec{q}_4)+\delta^{(3)}(\vec{p}_3-\vec{q}_4)\delta^{(3)}(\vec{p}_4-\vec{q}_1) \right)\left( \delta^{(3)}(\vec{p}_1-\vec{q}_2)\delta^{(3)}(\vec{p}_2-\vec{q}_3)+\delta^{(3)}(\vec{p}_2-\vec{q}_2)\delta^{(3)}(\vec{p}_1-\vec{q}_3) \right) \right. \nonumber \\
    &\left. \ \ \ \ \ \times g_1f_2f_3g_4 \right. \nonumber \\
        &\left. +\left( \delta^{(3)}(\vec{p}_3-\vec{q}_2)\delta^{(3)}(\vec{p}_4-\vec{q}_3) + \delta^{(3)}(\vec{p}_3-\vec{q}_3)\delta^{(3)}(\vec{p}_4-\vec{q}_2) \right)\left( \delta^{(3)}(\vec{p}_1-\vec{q}_1)\delta^{(3)}(\vec{p}_2-\vec{q}_4)+\delta^{(3)}(\vec{p}_2-\vec{q}_1)\delta^{(3)}(\vec{p}_1-\vec{q}_4) \right) \right. \nonumber \\
    &\left. \ \ \ \ \ \times f_1g_2g_3f_4 \right. \nonumber \\
        &\left. +\left( \delta^{(3)}(\vec{p}_3-\vec{q}_2)\delta^{(3)}(\vec{p}_4-\vec{q}_4)+\delta^{(3)}(\vec{p}_3-\vec{q}_4)\delta^{(3)}(\vec{p}_4-\vec{q}_2) \right)\left( \delta^{(3)}(\vec{p}_1-\vec{q}_1)\delta^{(3)}(\vec{p}_2-\vec{q}_3)+\delta^{(3)}(\vec{p}_2-\vec{q}_1)\delta^{(3)}(\vec{p}_1-\vec{q}_3) \right) \right. \nonumber \\
    &\left. \ \ \ \ \ \times f_1g_2f_3g_4 \right. \nonumber \\
        &\left. +\left( \delta^{(3)}(\vec{p}_3-\vec{q}_3)\delta^{(3)}(\vec{p}_4-\vec{q}_4)+\delta^{(3)}(\vec{p}_3-\vec{q}_4)\delta^{(3)}(\vec{p}_4-\vec{q}_3) \right)\left( \delta^{(3)}(\vec{p}_1-\vec{q}_1)\delta^{(3)}(\vec{p}_2-\vec{q}_2)+\delta^{(3)}(\vec{p}_2-\vec{q}_1)\delta^{(3)}(\vec{p}_1-\vec{q}_2) \right) \right. \nonumber \\
    &\left. \ \ \ \ \ \times f_1f_2g_3g_4  \right\}
    \end{align}
Inserting the explicit expressions for $f_i$ and $g_i$ from Eq.~(\ref{inoutwaves}), and writing the four-dimensional integration measure in cylindrical coordinates as $d^4x=dt\,dz\,d\varphi\, drr$, the integrations over the intermediate momenta can be performed using the three-dimensional Dirac delta functions, yielding
\begin{align}
    V_{A1}&=-i\lambda\int dt\,dz\,d\varphi \,dr\,r  \nonumber \\
        &\times \Bigg\{\Big[J_{l_{1}}(rp_{1\perp})J_{l_{2}}(rp_{2\perp})J_{l_{3}}(rp_{3\perp})J_{l_{4}}(rp_{4\perp})e^{-i(E_{p_1}^{(\sigma)}+E_{p_2}^{(\sigma)}-E_{p_3}^{(\sigma)}-E_{p_4}^{(\sigma)})t}e^{i(p_{1z}+p_{2z}-p_{3z}-p_{4z})z}e^{i(l_{1}+l_{2}-l_{3}-l_{4})\varphi} \nonumber \\
        &+J_{l_{1}}(rp_{1\perp})J_{l_{2}}(rp_{2\perp})J_{l_{3}}(rp_{3\perp})J_{l_{4}}(rp_{4\perp})e^{-i(E_{p_1}^{(\sigma)}-E_{p_2}^{(\sigma)}+E_{p_3}^{(\sigma)}-E_{p_4}^{(\sigma)})t}e^{i(p_{1z}-p_{2z}+p_{3z}-p_{4z})z}e^{i(l_{1}-l_{2}+l_{3}-l_{4})\varphi} \nonumber \\
		&+J_{l_{1}}(rp_{1\perp})J_{l_{2}}(rp_{2\perp})J_{l_{3}}(rp_{3\perp})J_{l_{4}}(rp_{4\perp})e^{-i(E_{p_1}^{(\sigma)}-E_{p_2}^{(\sigma)}-E_{p_3}^{(\sigma)}+E_{p_4}^{(\sigma)})t}e^{i(p_{1z}-p_{2z}-p_{3z}+p_{4z})z}e^{i(l_{1}-l_{2}-l_{3}+l_{4})\varphi} \nonumber \\
		&+J_{l_{1}}(rp_{1\perp})J_{l_{2}}(rp_{2\perp})J_{l_{3}}(rp_{3\perp})J_{l_{4}}(rp_{4\perp})e^{-i(-E_{p_1}^{(\sigma)}+E_{p_2}^{(\sigma)}+E_{p_3}^{(\sigma)}-E_{p_4}^{(\sigma)})t}e^{i(-p_{1z}+p_{2z}+p_{3z}-p_{4z})z}e^{i(-l_{1}+l_{2}+l_{3}-l_{4})\varphi} \nonumber \\
		&+J_{l_{1}}(rp_{1\perp})J_{l_{2}}(rp_{2\perp})J_{l_{3}}(rp_{3\perp})J_{l_{4}}(rp_{4\perp})e^{-i(-E_{p_1}^{(\sigma)}+E_{p_2}^{(\sigma)}-E_{p_3}^{(\sigma)}+E_{p_4}^{(\sigma)})t}e^{i(-p_{1z}+p_{2z}-p_{3z}+p_{4z})z}e^{i(-l_{1}+l_{2}-l_{3}+l_{4})\varphi} \nonumber \\
		&+J_{l_{1}}(rp_{1\perp})J_{l_{2}}(rp_{2\perp})J_{l_{3}}(rp_{3\perp})J_{l_{4}}(rp_{4\perp})e^{-i(-E_{p_1}^{(\sigma)}-E_{p_2}^{(\sigma)}+E_{p_3}^{(\sigma)}+E_{p_4}^{(\sigma)})t}e^{i(-p_{1z}-p_{2z}+p_{3z}+p_{4z})z}e^{i(-l_{1}-l_{2}+l_{3}+l_{4})\varphi}\Big] \Bigg\}.
\end{align}
Performing the integrations over $t$, $z$, and $\varphi$ produces Dirac delta functions associated with energy and longitudinal-momentum conservation, together with a Kronecker delta enforcing angular-momentum conservation along the rotation axis. The vertex function then becomes
\begin{align}
    V_{A1}&= -i\lambda\int \, dr\,r\nonumber \\
    &\times \Big[ \delta(E_{p_1}^{(\sigma)}+E_{p_2}^{(\sigma)}-E_{p_3}^{(\sigma)}-E_{p_4}^{(\sigma)})\delta(p_{1z}+p_{2z}-p_{3z}-p_{4z})\delta_{l_{1}+l_{2}, l_{3}+l_{4}}J_{l_{1}}(rp_{1\perp})J_{l_{2}}(rp_{2\perp})J_{l_{3}}(rp_{3\perp})J_{l_{4}}(rp_{4\perp}) \nonumber \\
		&+\delta(E_{p_1}^{(\sigma)}-E_{p_2}^{(\sigma)}+E_{p_3}^{(\sigma)}-E_{p_4}^{(\sigma)})\delta(p_{1z}-p_{2z}+p_{3z}-p_{4z})\delta_{l_{1}+l_{3}, l_{2}+l_{4}}J_{l_{1}}(rp_{1\perp})J_{l_{2}}(rp_{2\perp})J_{l_{3}}(rp_{3\perp})J_{l_{4}}(rp_{4\perp}) \nonumber \\
        &+\delta(E_{p_1}^{(\sigma)}-E_{p_2}^{(\sigma)}-E_{p_3}^{(\sigma)}+E_{p_4}^{(\sigma)})\delta(p_{1z}-p_{2z}-p_{3z}+p_{4z})\delta_{l_{1}+l_{4}, l_{2}+l_{3}}J_{l_{1}}(rp_{1\perp})J_{l_{2}}(rp_{2\perp})J_{l_{3}}(rp_{3\perp})J_{l_{4}}(rp_{4\perp}) \nonumber \\
		&+\delta(-E_{p_1}^{(\sigma)}+E_{p_2}^{(\sigma)}+E_{p_3}^{(\sigma)}-E_{p_4}^{(\sigma)})\delta(-p_{1z}+p_{2z}+p_{3z}-p_{4z})\delta_{l_{2}+l_{3}, l_{1}+l_{4}}J_{l_{1}}(rp_{1\perp})J_{l_{2}}(rp_{2\perp})J_{l_{3}}(rp_{3\perp})J_{l_{4}}(rp_{4\perp}) \nonumber \\
		&+\delta(-E_{p_1}^{(\sigma)}+E_{p_2}^{(\sigma)}-E_{p_3}^{(\sigma)}+E_{p_4}^{(\sigma)})\delta(-p_{1z}+p_{2z}-p_{3z}+p_{4z})\delta_{l_{2}+l_{4}, l_{1}+l_{3}}J_{l_{1}}(rp_{1\perp})J_{l_{2}}(rp_{2\perp})J_{l_{3}}(rp_{3\perp})J_{l_{4}}(rp_{4\perp}) \nonumber \\
		&+\delta(-E_{p_1}^{(\sigma)}-E_{p_2}^{(\sigma)}+E_{p_3}^{(\sigma)}+E_{p_4}^{(\sigma)})\delta(-p_{1z}-p_{2z}+p_{3z}+p_{4z})\delta_{l_{3}+l_{4}, l_{1}+l_{2}}J_{l_{1}}(rp_{1\perp})J_{l_{2}}(rp_{2\perp})J_{l_{3}}(rp_{3\perp})J_{l_{4}}(rp_{4\perp})\Big].
\end{align}  
\end{widetext}
The expression above contains the six possible assignments of two incoming and two outgoing external modes. Since the six contributions are related by permutations of the identical external sigma states, they give equivalent contributions to the vertex function. We may therefore write
\begin{align}
        V_{A1}&=-6i\lambda\delta \big(E_{p_1}^{(\sigma)}+E_{p_2}^{(\sigma)}-E_{p_3}^{(\sigma)}-E_{p_4}^{(\sigma)}\big)\nonumber \\
        &\times \delta(p_{1z}+p_{2z}-p_{3z}-p_{4z})\delta_{l_{1}+l_{2}, l_{3}+l_{4}} \nonumber \\ 
		&\times \int dr\,r\,J_{l_{1}}(rp_{1\perp})J_{l_{2}}(rp_{2\perp})J_{l_{3}}(rp_{3\perp})J_{l_{4}}(rp_{4\perp}).
        \label{vertexA1compact}
\end{align} 
Using the explicit expression for the sigma energy in Eq.~(\ref{EnergySigmaMeson}), Eq.~(\ref{vertexA1compact}) becomes
\begin{align}
    V_{A1}=&-6i\lambda
    \delta\big( \tilde{E}^{(\sigma)}_1+\tilde{E}^{(\sigma)}_2-\tilde{E}^{(\sigma)}_3-\tilde{E}^{(\sigma)}_4 \nonumber \\
        &-\Omega\left(l_1+l_2-l_3-l_4\right)\big)\nonumber\\
    &\times\delta\left(p_{3z}+p_{4z}-p_{1z}-p_{2z}\right)
    \delta_{l_1+l_2,l_3+l_4}\nonumber\\
    &\times\int dr\,r\,
    J_{l_1}(rp_{1\perp})
    J_{l_2}(rp_{2\perp})
    J_{l_3}(rp_{3\perp})
    J_{l_4}(rp_{4\perp}),
    \label{vertexA1final}
\end{align}
where we have defined 
\begin{equation}
\tilde{E}^{(\sigma)}_i=\sqrt{p^{2}_{i\perp}+p_{iz}^{2}+m_\sigma^2}
\end{equation}
with $i=1,2,3,4$.

Equation~(\ref{vertexA1final}) shows that the interaction no longer has the usual point-like structure in momentum space. Instead of the four-dimensional Dirac delta associated with a plane-wave basis, the vertex exhibits the conservation laws associated with the symmetries that remain manifest in the rotating system: the conserved quantity associated with time translations in the rotating frame, the longitudinal momentum, and the angular momentum along the rotation axis. The transverse dependence, by contrast, is contained in a non-trivial radial overlap integral. This structure reflects the loss of translational invariance in the transverse plane in the presence of global rotation.

The Bessel functions appearing in Eq.~(\ref{vertexA1final}) originate from the use of scalar modes adapted to the rotating background. These modes are not simply plane waves rewritten in cylindrical coordinates, but eigenfunctions of the corresponding kinetic operator. In this respect, they play a role analogous to that of the Ritus eigenfunctions~\cite{Ritus:1972ky} for charged particles in an external magnetic field, where the plane-wave basis is likewise replaced by eigenfunctions adapted to the background. The radial integral can therefore be interpreted as the overlap of four rotating modes, encoding the non-trivial transverse structure of the interaction.

The radial domain is further constrained by causality. A rigidly rotating system must remain entirely within the light cylinder; hence, we restrict the system to a cylinder of radius $R$ satisfying
\begin{equation}
0\leq r\leq R, \qquad \Omega R<1.
\label{causalitycondition}
\end{equation}
Accordingly, the radial overlap entering the interaction vertices is restricted to this causally allowed region.

We additionally impose a Dirichlet boundary condition at the surface of the cylinder,
\begin{equation}
\phi(r=R)=0,
\label{DirichletCondition}
\end{equation}
which confines the scalar modes to $0\leq r\leq R$. For the radial solutions, this condition requires
\begin{equation}
J_l(p_\perp R)=0,
\end{equation}
so that the transverse momentum becomes discrete,
\begin{equation}
p_{\perp,nl}=
\frac{\alpha_{l,n}}{R},
\label{TransverseMomentumQuantization}
\end{equation}
where $\alpha_{l,n}$ is the $n$th positive zero of $J_l$. The scalar modes are therefore labeled by the quantum numbers $(n,l,p_z)$, with
\begin{equation}
\tilde{E}^{(\sigma)}_{nl}=\sqrt{\frac{\alpha_{l,n}^{2}}{R^{2}}+p_z^{2}+m_\sigma^{2}}.
\label{DiscreteScalarEnergy}
\end{equation}
With these conditions, the four-sigma interaction vertex takes the final form
\begin{align}
    V_{A1}=&-6i\lambda
    \delta\big( \tilde{E}^{(\sigma)}_{n_1l_1}+\tilde{E}^{(\sigma)}_{n_2l_2}-\tilde{E}^{(\sigma)}_{n_3l_3}-\tilde{E}^{(\sigma)}_{n_4l_4} \nonumber \\
        &-\Omega\left(l_1+l_2-l_3-l_4\right)\big)\nonumber\\
    &\times\delta\left(p_{3z}+p_{4z}-p_{1z}-p_{2z}\right)
    \delta_{l_1+l_2,l_3+l_4}\nonumber\\
    &\times\int_0^R dr\,r\,\prod_{i=1}^{4}J_{l_i}\left(\frac{\alpha_{l_i,n_i}}{R}r\right).
    \label{VA1finalcausalDirichlet}
\end{align}
Thus, the causality condition determines the allowed radial extent of the rotating system, while the boundary condition leads to a discrete transverse spectrum and modifies the transverse structure of the interaction vertex.

We next consider the self-interaction of four neutral pion fields, corresponding to diagram $A2$ in Fig.~\ref{fig1}. The associated vertex function is
\begin{equation}
    V_{A2}=\langle\pi_{0p_{4}}\pi_{0p_{3}}|i\int d^{4}x\left(-\frac{\lambda\pi_0^{4}(x)}{4}\right)|\pi_{0p_{2}}\pi_{0p_{1}}\rangle.
    \label{vertexneutralpionsinitial}
\end{equation}
The derivation is identical to that of the four-sigma interaction, with the sigma mass replaced by the neutral pion mass in the external modes. The causality and boundary conditions introduced above apply without modification. Defining
\begin{equation}
\tilde{E}^{(\pi)}_{nl}=\sqrt{\frac{\alpha_{l,n}^2}{R^2}+p_z^2+m_\pi^2},
\label{DiscretePionEnergy}
\end{equation}
the resulting vertex function is
\begin{align}
V_{A2}=&-6i\lambda
\delta\bigg(\tilde{E}^{(\pi)}_{n_1l_1}+\tilde{E}^{(\pi)}_{n_2l_2}-\tilde{E}^{(\pi)}_{n_3l_3}-\tilde{E}^{(\pi)}_{n_4l_4}\nonumber\\
&-\Omega\left(l_1+l_2-l_3-l_4\right)\bigg)\nonumber\\
&\times \delta\left(p_{3z}+p_{4z}-p_{1z}-p_{2z}\right)\delta_{l_1+l_2,l_3+l_4}\nonumber\\
&\times \int_0^R dr\,r\,\prod_{i=1}^{4}J_{l_i}\left(\frac{\alpha_{l_i,n_i}}{R}r\right).
\label{vertexA2final}
\end{align}
As expected, Eq.~(\ref{vertexA2final}) has the same structure as the four-sigma vertex, with the corresponding change in the mass of the external modes.

We next consider the mixed four-meson interaction involving two sigma and two neutral pion fields, corresponding to diagram $A3$ in Fig.~\ref{fig1}. The associated vertex function is
\begin{equation}
    V_{A3}=
    \langle \sigma_{p_4}\sigma_{p_3}|
    i\int d^4x
    \left(
    -\frac{\lambda}{2}\sigma^2(x)\pi_0^2(x)
    \right)
    |\pi_{0p_2}\pi_{0p_1}\rangle.
    \label{VA3initial}
\end{equation}
The derivation follows directly from the previous cases, with the corresponding sigma and neutral pion energies determined by their respective masses. The resulting vertex function is
\begin{align}
V_{A3}&=-i2\lambda \delta\left(\tilde{E}_{n_1l_1}^{(\pi)}+\tilde{E}_{n_2l_2}^{(\pi)}-\tilde{E}_{n_3l_3}^{(\sigma)}-\tilde{E}_{n_4l_4}^{(\sigma)}\right.\nonumber \\
&\left.-\Omega\left(l_1+l_2-l_3-l_4\right)\right)\nonumber\\
&\times \delta(p_{3z}+p_{4z}-p_{1z}-p_{2z}) \delta_{l_1+l_2,l_3+l_4} \nonumber\\
&\times \int_0^R dr\,r\,\prod_{i=1}^{4}J_{l_i}\left(\frac{\alpha_{l_i,n_i}}{R}r\right) .
\label{vertexA3final}
\end{align}

We next consider the interaction among four charged pions, corresponding to diagram $A4$ in Fig.\ref{fig1}, with two positively and two negatively charged external states. Although the calculation follows the same general procedure as in the previous cases, the charged pion fields require a different mode expansion from that of the neutral scalar fields. The corresponding vertex function is
\begin{equation}
    V_{A4}=\langle\pi_{-p_{4}}\pi_{+p_{3}}|i\int d^{4}x\left(-\lambda\pi^2_+(x)\pi^2_-(x)\right)|\pi_{-p_{2}}\pi_{+p_{1}}\rangle.
    \label{V4initial}
\end{equation}
The one-particle states are defined as
\begin{align}
    |\pi_+\rangle&=\sqrt{2E^{(\pi)}_p}a_{\vec{p}}^\dagger|0\rangle, \nonumber \\
    |\pi_-\rangle&=\sqrt{2E^{(\pi)}_p}b_{\vec{p}}^\dagger|0\rangle,
    \label{ketschargedpions}
\end{align}
while the charged pion fields are expanded as
\begin{align}
    \pi_+&=\int\frac{d^3p}{(2\pi)^3}\frac{1}{\sqrt{2E^{(\pi)}_p}}\left( a_{\vec{p}}\, f(\vec{x},\vec{p})+b_{\vec{p}}^\dagger \,g(\vec{x},\vec{p}) \right), \nonumber \\
    \pi_-&=\int\frac{d^3p}{(2\pi)^3}\frac{1}{\sqrt{2E^{(\pi)}_p}}\left( b_{\vec{p}}\, f(\vec{x},\vec{p})+a_{\vec{p}}^\dagger \,g(\vec{x},\vec{p}) \right),
    \label{wavefunctionschargedpions}
\end{align}
with
\begin{equation}
    E_p^{(\pi)}=\sqrt{p_\perp^2+p_z^2+m_\pi^2}-\Omega l_p,
    \label{energypioncharged}
\end{equation}
where $f(\vec{x},\vec{p})$ and $g(\vec{x},\vec{p})$ are the rotating scalar modes defined in Eq.~(\ref{inoutwaves}). Here, $a_{\vec p}^{\dagger}$ ($a_{\vec p}$) and $b_{\vec p}^{\dagger}$ ($b_{\vec p}$) denote the creation (annihilation) operators for positively and negatively charged pions, respectively.

Substituting Eqs.~(\ref{ketschargedpions}) and~(\ref{wavefunctionschargedpions}) into Eq.~(\ref{V4initial}), we obtain
\begin{align}
    V_{A4}&=-i\lambda \int \frac{d^4x\,d^3q_1\,d^3q_2\,d^3q_3\,d^3q_4}{(2\pi)^{12} 4\sqrt{E_{q_1}^{(\pi)}E_{q_2}^{(\pi)}E_{q_3}^{(\pi)}E_{q_4}^{(\pi)}}}\nonumber \\
    &\times 4\sqrt{E_{p_1}^{(\pi)}E_{p_2}^{(\pi)}E_{p_3}^{(\pi)}E_{p_4}^{(\pi)}} \nonumber \\
    &\times \langle0|N\left[b_{\vec{p}_4}a_{\vec{p}_3}\left(a_{\vec{q}_4} \, f_4+ b^\dagger_{\vec{q}_4}\,g_4\right)\left( a_{\vec{q}_3} \, f_3+ b^\dagger_{\vec{q}_3}\,g_3 \right)\right. \nonumber \\
    &\left. \times \left( b_{\vec{q}_2}f_2+a^\dagger_{\vec{q}_2}g_2 \right)\left( b_{\vec{q}_1}f_1+a^\dagger_{\vec{q}_1}g_1 \right)b^\dagger_{\vec{p}_2}a^\dagger_{\vec{p}_1}\right]|0\rangle.
    \label{V4srtartingcomputation}
\end{align}
In contrast to the neutral scalar case, the charged fields contain independent particle and antiparticle operators. For the external-state assignment considered here, only those terms containing the appropriate $a$, $b$, $a^\dagger$, and $b^\dagger$ operators contribute to the matrix element. There are four equivalent contractions, corresponding to the $2!\,2!$ permutations of the two identical $\pi^+$ and two identical $\pi^-$ fields. Using the canonical commutation relations and performing the integrations over the intermediate momenta, the resulting combinatorial factor is therefore $4$. The vertex function becomes
\begin{align}
    V_{A4}&=-i4\lambda \delta \big(\tilde{E}_{n_1,l_1}^{(\pi)}+\tilde{E}_{n_2,l_2}^{(\pi)}-\tilde{E}_{n_3,l_3}^{(\pi)}-\tilde{E}_{n_4,l_4}^{(\pi)}\nonumber \\
    &-\Omega(l_1+l_2-l_3-l_4)\big)\nonumber \\
    &\times \delta(p_{1z}+p_{2z}-p_{3z}-p_{4z})\delta_{l_1+l_2,l_3+l_4}  \nonumber \\
    &\times \int_0^R dr\,r\,\prod_{i=1}^{4}J_{l_i}\left(\frac{\alpha_{l_i,n_i}}{R}r\right).
    \label{VA4final}
\end{align}

We next consider the mixed four-meson interaction involving two sigma fields and two charged pions, corresponding to diagram $A5$ in Fig.~\ref{fig1}. The associated vertex function is
\begin{equation}
    V_{A5}=\langle \sigma_{p_4} \sigma_{p_3} |i\int d^4x \big(-\lambda \sigma^2(x)\pi_+(x)\pi_-(x)\big) |\pi_{+p_2} \pi_{-p_1}\rangle.
    \label{VA5initial}
\end{equation}
This case is particularly useful because neutral and charged external fields are present simultaneously. Three distinct sets of creation and annihilation operators are therefore required, associated with the $\sigma$, $\pi_-$, and $\pi_+$ fields. We define the corresponding one-particle states as
\begin{align}
    |\sigma\rangle&=\sqrt{2E_p^{(\sigma)}}a^\dagger_p|0\rangle, \nonumber \\
    |\pi_-\rangle&=\sqrt{2E_p^{(\pi)}}b_p^\dagger|0\rangle, \nonumber \\
    |\pi_+\rangle&=\sqrt{2E_p^{(\pi)}}c_p^\dagger|0\rangle,
    \label{stateschargedsigma}
\end{align}
while the corresponding field expansions are
\begin{align}
    \sigma&=\int \frac{d^3p}{(2\pi)^3}\frac{1}{\sqrt{2E_p^{(\sigma)}}}\left[ a_{\vec{p}}f(\vec{x},\vec{p})+a^\dagger_{\vec{p}}\,g(\vec{x},\vec{p}) \right], \nonumber \\
    \pi_-&=\int \frac{d^3p}{(2\pi)^3}\frac{1}{\sqrt{2E_p^{(\pi)}}}\left[ b_{\vec{p}}f(\vec{x},\vec{p})+c^\dagger_{\vec{p}}\,g(\vec{x},\vec{p}) \right], \nonumber \\
    \pi_+&=\int \frac{d^3p}{(2\pi)^3}\frac{1}{\sqrt{2E_p^{(\pi)}}}\left[ c_{\vec{p}}f(\vec{x},\vec{p})+b^\dagger_{\vec{p}}\,g(\vec{x},\vec{p}) \right].
    \label{funcionschargedsigma}
\end{align}
Here, $E_p^{(\sigma)}$ and $E_p^{(\pi)}$ are given by Eqs.(\ref{EnergySigmaMeson}) and~(\ref{energypioncharged}), respectively, while the rotating scalar modes $f(\vec x,\vec p)$ and $g(\vec x,\vec p)$ are defined in Eq.~(\ref{inoutwaves}). The operators $a^\dagger_{\vec p}$ and $a_{\vec p}$ are associated with the sigma field, whereas $b^\dagger_{\vec p}$, $b_{\vec p}$ and $c^\dagger_{\vec p}$, $c_{\vec p}$ are associated with the negatively and positively charged pion sectors, respectively.

Substituting Eqs.~(\ref{stateschargedsigma}) and~(\ref{funcionschargedsigma}) into Eq.~(\ref{VA5initial}), we obtain
\begin{align}
   V_{A5}&=-i\lambda \int \frac{d^4x\,d^3q_1\,d^3q_2\,d^3q_3\,d^3q_4}{(2\pi)^{12}\sqrt{E_{q_1}^{(\pi)}E_{q_2}^{(\pi)}E_{q_3}^{(\sigma)}E_{q_4}^{(\sigma)}}} \nonumber \\
   &\times \sqrt{E_{p_1}^{(\pi)}E_{p_2}^{(\pi)}E_{p_3}^{(\sigma)}E_{p_4}^{(\sigma)}} \nonumber \\
   &\times \langle 0|N\left[ a_{\vec{p}_4}a_{\vec{p}_3}(a_{\vec{q}_4}f_4+a^\dagger_{\vec{q}_4}g_4)(a_{\vec{q}_3}f_3+a^\dagger_{\vec{q}_3}g_3)\right. \nonumber \\
   &\left. \times(b_{\vec{q}_2}f_2+c^\dagger_{\vec{q}_2}g_2)(c_{\vec{q}_1}f_1+b^\dagger_{\vec{q}_1}g_1)c^\dagger_{\vec{p}_2}b^\dagger_{\vec{p}_1} \right] |0\rangle,
   \label{Va5initialexplicit}
\end{align}
where $N$ denotes normal ordering and for compactness $f_i=f(\vec{x},\vec{q}_i)$ and $g_i=g(\vec{x},\vec{q}_i)$. 

For the external-state assignment considered here, only those terms containing the appropriate $a$, $b$, $c$, $a^\dagger$, $b^\dagger$, and $c^\dagger$ operators contribute to the matrix element. Expanding the product, using the canonical commutation relations, and performing the normal ordering, we obtain
\begin{widetext}
  \begin{align}
    V_{A5}
   &=-i\lambda \int d^4x\,d^3q_1\,d^3q_2\,d^3q_3\,d^3q_4 \sqrt{\frac{E_{p_1}^{(\pi)}E_{p_2}^{(\pi)}E_{p_3}^{(\sigma)}E_{p_4}^{(\sigma)}}{E_{q_1}^{(\pi)}E_{q_2}^{(\pi)}E_{q_3}^{(\sigma)}E_{q_4}^{(\sigma)}}}\left[ \delta^{(3)}(\vec{p}_4-\vec{q}_4)\delta^{(3)}(\vec{p}_3-\vec{q}_3)\delta^{(3)}(\vec{p}_2-\vec{q}_1)\delta^{(3)}(\vec{p}_1-\vec{q_2})\right. \nonumber \\
   &\left.+\delta^{(3)}(\vec{p}_3-\vec{q}_4)\delta^{(3)}(\vec{p}_4-\vec{q}_3)\delta^{(3)}(\vec{p}_2-\vec{q}_1)\delta^{(3)}(\vec{p}_1-\vec{q_2})\right]f_4f_3g_2g_1.
   \label{VA5afternormalordering}
\end{align}  
The two terms in Eq.~(\ref{VA5afternormalordering}) correspond to the two equivalent assignments of the identical outgoing sigma fields. Inserting the explicit expressions for the rotating modes from Eq.~(\ref{inoutwaves}), we obtain
\begin{align}
   V_{A5}&= -i\lambda \int d^4x\,d^3q_1\,d^3q_2\,d^3q_3\,d^3q_4\sqrt{\frac{E_{p_1}^{(\pi)}E_{p_2}^{(\pi)}E_{p_3}^{(\sigma)}E_{p_4}^{(\sigma)}}  {E_{q_1}^{(\pi)}E_{q_2}^{(\pi)}E_{q_3}^{(\sigma)}E_{q_4}^{(\sigma)}}}\left[ \delta^{(3)}(\vec{p}_4-\vec{q}_4)\delta^{(3)}(\vec{p}_3-\vec{q}_3)\delta^{(3)}(\vec{p}_2-\vec{q}_1)\delta^{(3)}(\vec{p}_1-\vec{q_2})\right. \nonumber \\
   &\left.+\delta^{(3)}(\vec{p}_3-\vec{q}_4)\delta^{(3)}(\vec{p}_4-\vec{q}_3)\delta^{(3)}(\vec{p}_2-\vec{q}_1)\delta^{(3)}(\vec{p}_1-\vec{q_2})\right]J_{l_{1}}(rq_{1\perp})J_{l_{2}}(rq_{2\perp})J_{l_{3}}(rq_{3\perp})J_{l_{4}}(rq_{4\perp})\nonumber \\
   &\times e^{-i(E_{q_4}^{(\sigma)}+E_{q_3}^{(\sigma)}-E_{q_2}^{(\pi)}-E_{q_1}^{(\pi)})t}e^{i(q_{4z}+q_{3z}-q_{2z}-q_{1z})z}e^{i(l_{4}+l_{3}-l_{2}-l_{1})\varphi}.
   \label{VA5explicitfunctions}
\end{align}
\end{widetext}
The integrations over the intermediate momenta can then be performed using the three-dimensional Dirac delta functions. Since the two contributions differ only by the interchange of the identical outgoing sigma fields, they give identical results, yielding an overall factor of $2$. We thus obtain
\begin{align}
   V_{A5}&= -i2\lambda \int dt\,dz\,d\varphi\,drr \, e^{-i(E_{p_4}^{(\sigma)}+E_{p_3}^{(\sigma)}-E_{p_2}^{(\pi)}-E_{p_1}^{(\pi)})t}\nonumber \\
   &\times e^{i(p_{4z}+p_{3z}-p_{2z}-p_{1z})z}e^{i(l_{4}+l_{3}-l_{2}-l_{1})\varphi}\nonumber \\
   &\times J_{l_{1}}(rp_{1\perp})J_{l_{2}}(rp_{2\perp})J_{l_{3}}(rp_{3\perp})J_{l_{4}}(rp_{4\perp}).
   \label{VA5aftermomentaint}
\end{align}
Performing the integrations over $t$, $z$, and $\varphi$ yields the corresponding conservation conditions for energy, longitudinal momentum, and angular momentum along the rotation axis. The vertex function then becomes
\begin{align}
   V_{A5}&=-i2\lambda \delta \big(\tilde{E}_4^{(\sigma)}+\tilde{E}_3^{(\sigma)}-\tilde{E}_2^{(\pi)}-\tilde{E}_1^{(\pi)} \nonumber \\
   &-\Omega(l_4+l_3-l_2-l_1)\big) \nonumber \\
   &\times \delta(p_{4z}+p_{3z}-p_{2z}-p_{1z})\delta_{l_3+l_4,l_1+l_2} \nonumber \\
   &\times \int dr r J_{l_{1}}(rp_{1\perp})J_{l_{2}}(rp_{2\perp})J_{l_{3}}(rp_{3\perp})J_{l_{4}}(rp_{4\perp}),
   \label{VA5afterposicionint}
\end{align}
with $\tilde{E}_i^{(\pi/\sigma)}=\sqrt{p_{i\perp}^2+p_{iz}^2+m_{\pi/\sigma}^2}$. 

Finally, imposing the causality restriction and the Dirichlet boundary condition introduced above, the transverse momenta are restricted to the discrete values given by Eq.~(\ref{TransverseMomentumQuantization}). The vertex function takes the final form
\begin{align}
    V_{A5}&=-i2\lambda \delta \big(\tilde{E}_{n_4l_4}^{(\sigma)}+\tilde{E}_{n_3l_3}^{(\sigma)}-\tilde{E}_{n_2l_2}^{(\pi)}-\tilde{E}_{n_1l_1}^{(\pi)} \nonumber \\
   &-\Omega(l_4+l_3-l_2-l_1)\big) \nonumber \\
   &\times \delta(p_{4z}+p_{3z}-p_{2z}-p_{1z})\delta_{l_3+l_4,l_1+l_2} \nonumber \\
   &\times \int_0^R dr\,r\,\prod_{i=1}^{4}J_{l_i}\left(\frac{\alpha_{l_i,n_i}}{R}r\right).
   \label{VA5final}
\end{align}

Finally, we consider the last four-meson interaction, involving two neutral and two charged pion fields and corresponding to diagram $A6$ in Fig.~\ref{fig1}. The associated vertex function is
\begin{equation}
    V_{A6}=\langle \pi_{0p_4} \pi_{0p_3} |i\int d^4x \big(-\lambda \pi_0^2(x)\pi_+(x)\pi_-(x)\big) |\pi_{+p_2} \pi_{-p_1}\rangle.
    \label{VA6initial}
\end{equation}
The derivation proceeds exactly as for $A5$, since the operator structure is unchanged. The only modification is the replacement of the two sigma fields by neutral pion fields, $\sigma\rightarrow\pi_0$, together with the corresponding change in the energies of the outgoing modes. The resulting vertex function is therefore
\begin{align}
    V_{A6}&=-i2\lambda \delta \big(\tilde{E}_{n_4l_4}^{(\pi)}+\tilde{E}_{n_3l_3}^{(\pi)}-\tilde{E}_{n_2l_2}^{(\pi)}-\tilde{E}_{n_1l_1}^{(\pi)} \nonumber \\
   &-\Omega(l_4+l_3-l_2-l_1)\big) \nonumber \\
   &\times \delta(p_{4z}+p_{3z}-p_{2z}-p_{1z})\delta_{l_3+l_4,l_1+l_2} \nonumber \\
   &\times \int_0^R dr\,r\,\prod_{i=1}^{4}J_{l_i}\left(\frac{\alpha_{l_i,n_i}}{R}r\right).
   \label{VA6final}
\end{align}

Having obtained the six four-meson vertices, we now turn to the interactions involving three mesonic fields.

\begin{figure}
    \centering
    \includegraphics[scale=0.6]{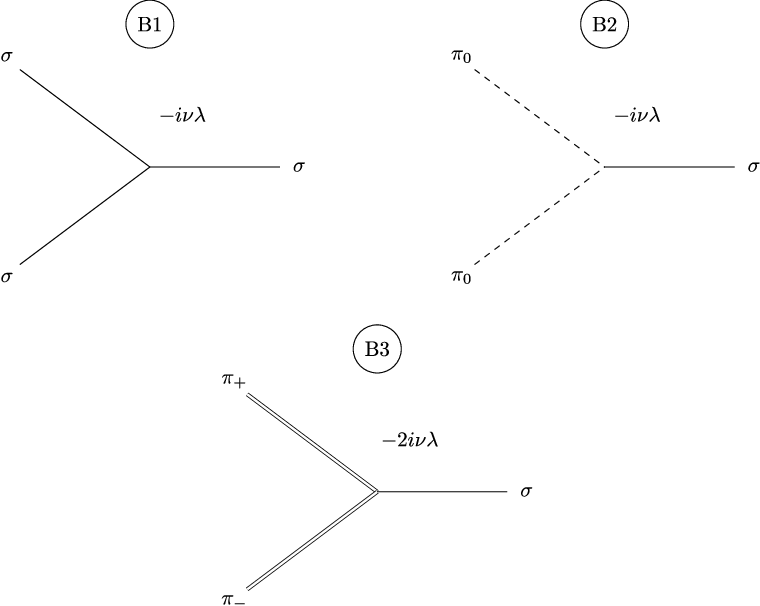}
    \caption{Feynman diagrams corresponding to the vertices with three meson fields interacting. Solid, dashed and double lines denote the sigma meson, the neutral pion and the charged pions, respectively.}
    \label{fig2}
\end{figure}

The three-meson interactions are represented by the diagrams shown in Fig.\ref{fig2}, labeled $Bi$, with $i=1,2,3$. We first consider $B1$, which describes the self-interaction of three sigma fields. From the corresponding term in the interaction Lagrangian density, Eq.~(\ref{interactinglagrangian}), the vertex function is
\begin{align}
	V_{B1}=\langle\sigma_{p_3}|i\int d^{4}x \left( -\nu \lambda \sigma^3(x) \right)|\sigma_{p_2}\sigma_{p_1}\rangle.
    \label{B1initial}
\end{align}
The one-particle states and the sigma-field expansion are given in Eqs.~(\ref{ketsfromvacuum}) and~(\ref{spectralrepsigma}), respectively, with the corresponding energy and rotating modes defined in Eqs.~(\ref{EnergySigmaMeson}) and~(\ref{inoutwaves}). Substituting these expressions into Eq.~(\ref{B1initial}) yields
\begin{align}
	V_{B1}&=-i\nu\lambda \int \frac{d^4x\,d^3q_1\,d^3q_2\,d^3q_3}{(2\pi)^{9}\sqrt{8E_{q_1}^{(\sigma)}E_{q_2}^{(\sigma)}E_{q_3}^{(\sigma)}}} \nonumber \\
    &\times \sqrt{8E_{p_1}^{(\sigma)}E_{p_2}^{(\sigma)}E_{p_3}^{(\sigma)}} \nonumber \\
    &\times \langle 0| N \left[ a_{\vec{p}_3}\big(a_{\vec{q}_1}f_1+a^{\dagger}_{\vec{q}_1}g_1\big)\right. \nonumber \\
    &\left. \times \big(a_{\vec{q}_2}f_2+a^{\dagger}_{\vec{q}_2}g_2\big)\big(a_{\vec{q}_3}f_3+a^{\dagger}_{\vec{q}_3}g_3\big)a_{\vec{p}_2}^{\dagger}a_{\vec{p}_1}^{\dagger}\right]|0\rangle
    \label{B1explicit},
\end{align}
where $N$ denotes normal ordering and, as before, $f_i$ and $g_i$ denote the rotating scalar modes in compact notation. Expanding the product and using the canonical commutation relations gives
\begin{widetext}
\begin{align}
    V_{B1}&=-i\nu\lambda\int \frac{d^4x\,d^3q_1\,d^3q_2\,d^3q_3}{(2\pi)^{9}}\sqrt{\frac{E_{p_1}^{(\sigma)}E_{p_2}^{(\sigma)}E_{p_3}^{(\sigma)}}{E_{q_1}^{(\sigma)}E_{q_2}^{(\sigma)}E_{q_3}^{(\sigma)}}} (2\pi)^{9} \nonumber \\ &
    \times \left\{ \delta^{(3)}(\vec{p}_3-\vec{q}_3)(\delta^{(3)}(\vec{p}_2-\vec{q}_1)\delta^{(3)}(\vec{p}_1-\vec{q}_2)+\delta^{(3)}(\vec{p}_1-\vec{q}_1)\delta^{(3)}(\vec{p}_2-\vec{q}_2))f_1f_2g_3 \right. \nonumber \\
    &\left. + \delta^{(3)}(\vec{p}_3-\vec{q}_2)(\delta^{(3)}(\vec{p}_2-\vec{q}_1)\delta^{(3)}(\vec{p}_1-\vec{q}_3)+\delta^{(3)}(\vec{p}_1-\vec{q}_1)\delta^{(3)}(\vec{p}_2-\vec{q}_3))f_1g_2f_3 \right. \nonumber \\
    &\left. + \delta^{(3)}(\vec{p}_3-\vec{q}_1)(\delta^{(3)}(\vec{p}_2-\vec{q}_2)\delta^{(3)}(\vec{p}_1-\vec{q}_3)+\delta^{(3)}(\vec{p}_1-\vec{q}_2)\delta^{(3)}(\vec{p}_2-\vec{q}_3))g_1f_2f_3 \right\}.
    \label{B1afternormalordering}
\end{align}
Using the three-dimensional Dirac delta functions to perform the integrations over the internal momenta and inserting the explicit rotating modes from Eq.~(\ref{inoutwaves}), we obtain
\begin{align}
    V_{B1}=-i6\nu\lambda \int dt\,dz\,d\varphi\,dr \,r J_{l_1}(rp_{1\perp})J_{l_2}(rp_{2\perp})J_{l_3}(rp_{3\perp})e^{-i(E_{p_1}^{(\sigma)}+E_{p_2}^{(\sigma)}-E_{p_3}^{(\sigma)})t}e^{i(p_{1z}+p_{2z}-p_{3z})z}e^{i(l_{1}+l_{2}-l_{3})\varphi}.
        \label{B1aftermomentaint}
\end{align}
\end{widetext}
For three identical sigma fields, there are $3!=6$ equivalent assignments connecting the external states. The integrations over $t$, $z$, and $\varphi$ enforce energy, longitudinal-momentum, and angular-momentum conservation, respectively, leading to
\begin{align}
    V_{B1}&=-i6\nu\lambda\int dr\,r J_{l_1}(rp_{1\perp})J_{l_2}(rp_{2\perp})J_{l_3}(rp_{3\perp}) \nonumber \\
    &\times \delta \left(\tilde{E}_{p_1}^{(\sigma)}+\tilde{E}_{p_2}^{(\sigma)}-\tilde{E}_{p_3}^{(\sigma)}-\Omega(l_1+l_2-l_3)\right) \nonumber \\
    &\times \delta(p_{1z}+p_{2z}-p_{3z})\delta_{l_1+l_2,l_3},
    \label{B1beforercausality}
\end{align}
where $\tilde{E}i^{(\sigma)}=\sqrt{p{i\perp}^2+p_{iz}^2+m_\sigma^2}$. Imposing the causality restriction and the Dirichlet boundary condition introduced above, gives the final form
\begin{align}
    V_{B1}&=-6i\nu \lambda \delta(p_{1z}+p_{2z}-p_{3z})\delta_{l_1+l_2,l_3} \nonumber \\
    &\times \delta \left(\tilde{E}_{n_1l_1}^{(\sigma)}+\tilde{E}_{n_2l_2}^{(\sigma)}-\tilde{E}_{n_3l_3}^{(\sigma)}-\Omega(l_1+l_2-l_3)\right) \nonumber \\
    &\times \int_0^R dr\, r \prod_{i=1}^{3}J_{l_i}\left(\frac{\alpha_{l_i,n_i}}{R}r\right).
    \label{B1final}
\end{align}

We next consider the interaction involving two neutral pions and one sigma field, corresponding to diagram $B2$ in Fig.~\ref{fig2}. The associated vertex function is
\begin{align}
	V_{B2}=\langle\sigma_{p_3}|i\int d^{4}x\left(-\nu \lambda\sigma(x)\pi_0^2(x)\right)|\pi_{0p_2}\pi_{0p_1}\rangle.
    \label{B2initial}
\end{align}
Unlike $B1$, this interaction involves two different mesonic species. Since the sigma field is uniquely associated with the outgoing state, the only equivalent contributions arise from interchanging the two identical incoming neutral pions. The corresponding combinatorial factor is therefore $2$. After normal ordering and performing the integrations over the internal momenta, we obtain
\begin{align}
    V_{B2}&=-i2\nu\lambda \int dt\,dz\,d\varphi\,dr\,r J_{l_1}(rp_{1\perp})J_{l_2}(rp_{2\perp})J_{l_3}(rp_{3\perp})\nonumber \\
    &\times e^{-i(E_{p_1}^{(\pi)}+E_{p_2}^{(\pi)}-E_{p_3}^{(\sigma)})t}e^{i(p_{1z}+p_{2z}-p_{3z})z}e^{i(l_{1}+l_{2}-l_{3})\varphi}
    \label{B2aftermomentaint}
\end{align}
Performing the integrations over $t$, $z$, and $\varphi$, and applying the causality and Dirichlet boundary conditions introduced above, yields
\begin{align}
    V_{B2}&=-2i\nu\lambda \delta(p_{1z}+p_{2z}-p_{3z})\delta_{l_1+l_2,l_3} \nonumber \\
    &\times \delta \left(\tilde{E}_{n_1l_1}^{(\pi)}+\tilde{E}_{n_2l_2}^{(\pi)}-\tilde{E}_{n_3l_3}^{(\sigma)}-\Omega(l_1+l_2-l_3)\right) \nonumber \\
    &\int_0^R dr\, r \prod_{i=1}^{3}J_{l_i}\left(\frac{\alpha_{l_i,n_i}}{R}r\right).
    \label{B2final}
\end{align}

The third and last interaction with three scalar fields corresponds to one sigma and two charged pions, corresponding to diagram $B3$ in Fig.~\ref{fig2}. The associated vertex function is
\begin{equation}
    V_{B3}=\langle\sigma_{p_3}|i\int d^{4}x\left(-2\nu\lambda\sigma(x)\pi_+(x)\pi_-(x)\right)|\pi_{+p_2}\pi_{-p_1}\rangle.
    \label{B3initial}
\end{equation}
This interaction involves the same three mesonic species introduced in $A5$, namely the sigma and the positively and negatively charged pion fields. We therefore use the one-particle states and field expansions defined in Eqs.~(\ref{stateschargedsigma}) and~(\ref{funcionschargedsigma}), respectively. Substituting them into Eq.~(\ref{B3initial}), we obtain
\begin{align}
   V_{B3}&=-i2\nu\lambda \int \frac{d^4x\,d^3q_1\,d^3q_2\,d^3q_3}{(2\pi)^9\sqrt{E_{q_1}^{(\pi)}E_{q_2}^{(\pi)}E_{q_3}^{(\sigma)}}} \nonumber \\
   &\times \sqrt{E_{p_1}^{(\pi)}E_{p_2}^{(\pi)}E_{p_3}^{(\sigma)}} \nonumber \\
   &\times \langle 0|N\left[ a_{\vec{p}_3}(a_{\vec{q}_3}f_3+a^\dagger_{\vec{q}_3}g_3)\right. \nonumber \\
   &\left. \times(b_{\vec{q}_2}f_2+c^\dagger_{\vec{q}_2}g_2)(c_{\vec{q}_1}f_1+b^\dagger_{\vec{q}_1}g_1)c^\dagger_{\vec{p}_2}b^\dagger_{\vec{p}_1} \right] |0\rangle,
   \label{B3initialexplicit}
\end{align}
where $N$ denotes normal ordering and, as before, $f_i$ and $g_i$ denote the rotating scalar modes in compact notation. Since the two incoming particles are distinguishable, there is a single operator assignment connecting the external states. Using the canonical commutation relations and performing the normal ordering then gives
\begin{widetext}
  \begin{equation}
    V_{B3}=-i2\nu\lambda \int d^4x\,d^3q_1\,d^3q_2\,d^3q_3  \sqrt{\frac{E_{p_1}^{(\pi)}E_{p_2}^{(\pi)}E_{p_3}^{(\sigma)}}  {E_{q_1}^{(\pi)}E_{q_2}^{(\pi)}E_{q_3}^{(\sigma)}}} \left[ \delta^{(3)}(\vec{p}_3-\vec{q}_3)\delta^{(3)}(\vec{p}_1-\vec{q}_2)\delta^{(3)}(\vec{p}_2-\vec{q}_1) \right] f_1f_2g_3.
    \label{B3afternormalordering}
\end{equation} 
Using the three-dimensional Dirac delta functions to perform the integrations over the internal momenta and inserting the explicit rotating modes from Eq.~(\ref{inoutwaves}), we find
\begin{align}
    V_{B3}=-i2\nu\lambda \int dt\,dz\,d\varphi\,dr\,r  J_{l_{1}}(rp_{1\perp})J_{l_{2}}(rp_{2\perp})J_{l_{3}}(rp_{3\perp}) e^{-i(E_{p_1}^{(\pi)}+E_{p_2}^{(\pi)}-E_{p_3}^{(\sigma)})t}e^{i(p_{1z}+p_{2z}-p_{3z})z}e^{i(l_{1}+l_{2}-l_{3})\varphi}
\end{align}
\end{widetext}
The integrations over $t$, $z$, and $\varphi$ enforce the same conservation conditions encountered in the previous three-meson vertices. Applying the causality restriction and Dirichlet boundary condition, the vertex function becomes
\begin{align}
    V_{B3}&=-2i\nu \lambda \delta(p_{1z}+p_{2z}-p_{3z})\delta_{l_1+l_2,l_3} \nonumber \\
    &\times \delta \left(\tilde{E}_{n_1l_1}^{(\pi)}+\tilde{E}_{n_2l_2}^{(\pi)}-\tilde{E}_{n_3l_3}^{(\sigma)}-\Omega(l_1+l_2-l_3)\right) \nonumber \\
    &\int_0^R dr\, r \prod_{i=1}^{3}J_{l_i}\left(\frac{\alpha_{l_i,n_i}}{R}r\right).
    \label{B3final}
\end{align}
It is worth emphasizing that, unlike in $B2$, the factor of $2$ in $B3$ does not originate from equivalent permutations of the external states. Since $\pi^+$ and $\pi^-$ are distinguishable, there is only one contributing operator assignment. Instead, this factor follows directly from the coefficient of the interaction term $-2\nu\lambda\sigma\pi_+\pi_-$ in the Lagrangian density.

\begin{figure}
    \centering
    \includegraphics[scale=0.6]{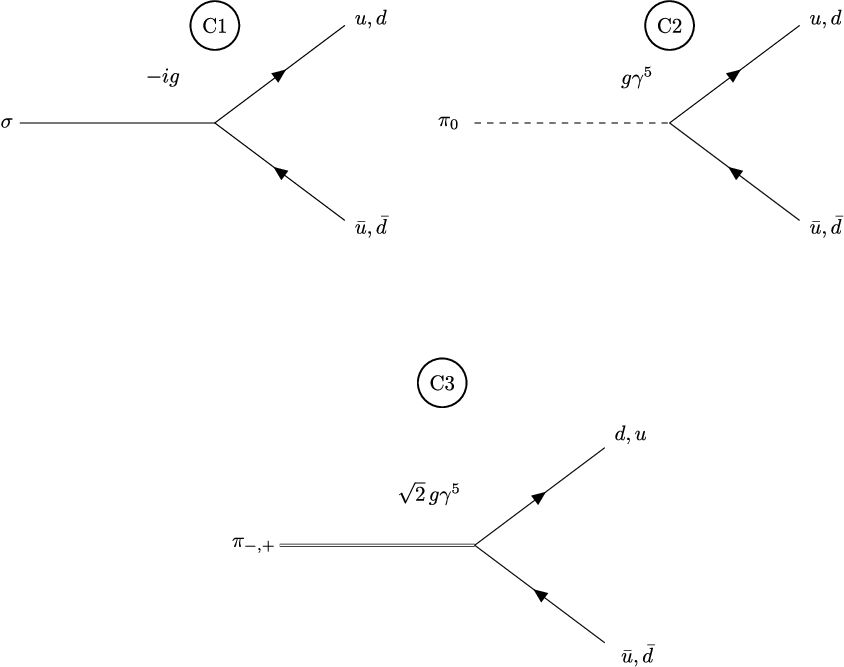}
    \caption{Feynman diagrams corresponding to the Yukawa interaction vertices involving meson and quarks fields. Solid, dashed, double lines and solid lines with arrows denote the sigma meson, the neutral pion, the charged pions and the quarks, respectively.}
    \label{fig3}
\end{figure}

We finally consider the Yukawa interactions between the mesonic and quark fields, represented by the three diagrams labeled $C1$, $C2$, and $C3$ in Fig.\ref{fig3}. In contrast to the purely mesonic vertices discussed above, these interactions involve fermionic modes in the rotating background.

We begin with the interaction between the sigma field and the quarks, corresponding to diagram $C1$ in Fig.\ref{fig3}. From the Yukawa interaction term in Eq.~(\ref{interactinglagrangian}), the associated vertex function can be written as
\begin{equation}
V_{C1}=\langle \bar{\psi}_{p_3,s_3}\psi_{p_2,s_2}| i\int d^4x
\left(-g\bar{\psi}(x)\sigma(x)\psi(x)\right)|\sigma_{p_1}\rangle ,
\label{C1initial}
\end{equation}
Unlike the purely mesonic vertices considered above, this interaction involves both positive- and negative-energy fermionic modes. To evaluate Eq.~(\ref{C1initial}), we first introduce the corresponding one-particle states as 
\begin{align}
|\psi_{\vec p,s}\rangle&=\sqrt{2E_p}\,b_{\vec p,s}^\dagger|0\rangle, \nonumber\\
|\bar\psi_{\vec p,s}\rangle&=\sqrt{2E_p}\,d_{\vec p,s}^\dagger|0\rangle,
\label{fermionstates}
\end{align}
where the index $s=1,2$ labels the two possible spin projections of the fermionic states. The fermionic field operator can be expanded in terms of positive- and negative-energy solutions of the Dirac equation as
\begin{align}
\psi(x)=\sum_s \int\frac{d^3p}{(2\pi)^3}\frac{1}{\sqrt{2E_p}}
\left[ b_{\vec{p},s}U(\vec{p})+d^\dagger_{\vec{p},s}V(\vec{p}) \right],
\label{fermionspectralrepresentation}
\end{align}
while the corresponding Dirac-adjoint field is
\begin{align}
\bar\psi(x)=\sum_s \int\frac{d^3p}{(2\pi)^3}\frac{1}{\sqrt{2E_p}}
\left[b^\dagger_{\vec p,s}\bar U(\vec p)+d_{\vec p,s}\bar V(\vec p)\right],
\label{adjointfermionspectralrepresentation}
\end{align}
where
\begin{equation}
\bar U(\vec p)=U^\dagger(\vec p)\gamma^0, \qquad \bar V(\vec p)=V^\dagger(\vec p)\gamma^0.
\end{equation}
The operators $b$, $b^\dagger$ and $d$, $d^\dagger$ annihilate and create quark and antiquark states, respectively, and satisfy the canonical anticommutation relations
\begin{align}
\left\{ b_{\vec p,s},b^\dagger_{\vec p\,',s'} \right\}&=(2\pi)^3\delta^{(3)}(\vec p-\vec p\,')\delta_{ss'},
\nonumber \\
\left\{ d_{\vec p,s},d^\dagger_{\vec p\,',s'} \right\}&=(2\pi)^3\delta^{(3)}(\vec p-\vec p\,')\delta_{ss'},
\label{fermionanticommutation}
\end{align}
with all other anticommutators vanishing.

For later convenience, we separate the temporal and longitudinal dependence of the positive- and negative-energy modes according to
\begin{align}
U(\vec p) &=e^{-iE_pt+ip_z z}\,u_p(r,\varphi), \nonumber \\
V(\vec p)&=e^{iE_pt-ip_z z}\,v_{p}(r,\varphi),
\label{UVfermionmode}
\end{align}
where the transverse and azimuthal dependence is contained in the spinors $u_p(r,\varphi)$ and $v_{p}(r,\varphi)$, respectively. 

The two spin projections are already incorporated into the fermionic modes used here. In writing the spinor solutions, the angular-momentum index associated with the $S_z=-1/2$ components has been shifted so that both spin projections can be expressed in terms of a single sum over $l$. Consequently, the components proportional to $J_l(p_\perp r)e^{il\varphi}$ carry $S_z=+1/2$, whereas those proportional to $J_{l+1}(p_\perp r)e^{i(l+1)\varphi}$ carry $S_z=-1/2$. Both contributions therefore have the same total angular-momentum projection,
\begin{equation}
j_z=l+\frac{1}{2}=(l+1)-\frac{1}{2}.
\end{equation}
Accordingly, no additional spin label is required in the notation adopted below. Then, their explicit forms are
\begin{align}
u_p(r,\varphi)=
\begin{pmatrix}
\big[\tilde{E}_p+m_f-p_z+ip_\perp\big]J_l(p_\perp r)
\\
\big[ \tilde{E}_p+m_f+p_z-ip_\perp\big]e^{i\varphi}J_{l+1}(p_\perp r)
\\
\big[- \tilde{E}_p+m_f+p_z-ip_\perp\big]J_l(p_\perp r)
\\
\big[- \tilde{E}_p+m_f-p_z+ip_\perp\big]e^{i\varphi}J_{l+1}(p_\perp r)
\end{pmatrix}
e^{il\varphi},
\label{upspinor}
\end{align}
and
\begin{align}
v_p(r,\varphi)=
\begin{pmatrix}
\big[ -\tilde{E}_p+m_f-p_z-ip_\perp\big]e^{-i\varphi}J_{l+1}(p_\perp r)
\\
\big[ \tilde{E}_p-m_f-p_z-ip_\perp\big]J_l(p_\perp r)
\\
\big[- \tilde{E}_p-m_f-p_z-ip_\perp\big]e^{-i\varphi}J_{l+1}(p_\perp r)
\\
\big[\tilde{E}_p+m_f-p_z-ip_\perp\big]J_l(p_\perp r)
\end{pmatrix}
e^{-il\varphi}.
\label{vmspinor}
\end{align}
Here, $\tilde E_p=\sqrt{p_\perp^2+p_z^2+m_f^2}$ corresponds to the vacuum dispersion relation, whereas $E_p=\tilde E_p-\Omega\left(l+\frac{1}{2}\right)$ is the corresponding energy in the rotating background. The rotational shift arises from the coupling of the angular velocity to the total angular momentum along the rotation axis, $J_z=L_z+S_z$.

Using the spectral representation of the sigma field given in Eq.~(\ref{spectralrepsigma}), together with Eqs.~(\ref{fermionspectralrepresentation}) and~(\ref{adjointfermionspectralrepresentation}) for the fermionic fields, Eq.~(\ref{C1initial}) becomes
\begin{align}
V_{C1}=&-ig\int\frac{d^4x \, d^3q_1 \, d^3q_2 \, d^3q_3}{(2\pi)^9\sqrt{8E_{q_1}^{(\sigma)}E_{q_2}^{(\psi)}E_{q_3}^{(\bar{\psi})}}} \nonumber \\
&\times \sqrt{8E_{p_1}^{(\sigma)}E_{p_2}^{(\psi)}E_{p_3}^{(\bar{\psi})}} \nonumber \\
&\times \langle 0|N \left[ d_{\vec{p}_3,s_3}b_{\vec{p}_2,s_2} \left(b^\dagger_{\vec{q}_2,s'_2}\bar U(\vec q_2)+d_{\vec{q}_2,s'_2}\bar V(\vec q_2)
\right) \right. \nonumber\\
&\left. \times \left( a_{\vec q_1}f_1+a^\dagger_{\vec q_1}g_1 \right) \right. \nonumber\\
&\left. \times \left(b_{\vec{q}_3,s'_3}U(\vec q_3) +d^\dagger_{\vec{q}_3,s'_3}V(\vec q_3)
\right) \right] a^\dagger_{\vec{p}_1}|0\rangle.
\label{C1explicit}
\end{align}

For the external-state assignment considered here, the normal ordering shows that only one operator combination contributes to the matrix element, namely $N\left[d_{\vec{p}_3,s_3}b_{\vec{p}_2,s_2}b^\dagger_{\vec{q}_2,s'_2}a_{\vec q_1}d^\dagger_{\vec{q}_3,s'_3}a^\dagger_{\vec p_1}\right]\bar U(\vec q_2)f_1V(\vec q_3)$. Using the canonical commutation and anticommutation relations, we obtain
\begin{align}
&\langle0|d_{\vec{p}_3,s_3}b_{\vec{p}_2,s_2}b^\dagger_{\vec{q}_2,s'_2}a_{\vec q_1}d^\dagger_{\vec{q}_3,s'_3}a^\dagger_{\vec p_1}|0\rangle \nonumber\\
&\,=(2\pi)^9 \delta^{(3)}(\vec q_1-\vec p_1) \delta^{(3)}(\vec q_2-\vec p_2)\delta_{s'_2,s_2} \delta^{(3)}(\vec q_3-\vec p_3)\delta_{s'_3,s_3}.
\label{C1contractions}
\end{align}
The integrations over $\vec q_1$, $\vec q_2$, and $\vec q_3$ can therefore be performed directly. After cancellation of the normalization factors, the vertex function reduces to
\begin{equation}
V_{C1}=-ig\int d^4x\, \bar U(\vec p_2)f_1V(\vec p_3).
\label{C1reduced}
\end{equation}
Using the explicit temporal and longitudinal dependence of the fermionic modes in Eq.~(\ref{UVfermionmode}), together with that of the sigma mode, the integrations over $t$ and $z$ yield
\begin{align}
V_{C1}&=-ig\, \delta\left( E_{p_1}^{(\sigma)}-E_{p_2}^{(\psi)}-E_{p_3}^{(\bar\psi)}\right)
\delta\left(p_{1z}-p_{2z}-p_{3z}\right) \nonumber \\
&\times \int_0^R dr\,r \int_0^{2\pi}d\varphi\,\bar u_{p_2}(r,\varphi)\,
J_{l_1}(p_{1\perp}r)e^{il_1\varphi}\,v_{p_3}(r,\varphi).
\label{C1tzintegrated}
\end{align}
Substituting the explicit forms of the spinors from Eqs.(\ref{upspinor}) and(\ref{vmspinor}), we obtain
\begin{widetext}
\begin{align}
    V_{C1}&=-ig\, \delta\left( E_{p_1}^{(\sigma)}-E_{p_2}^{(\psi)}-E_{p_3}^{(\bar\psi)}\right)
\delta\left(p_{1z}-p_{2z}-p_{3z}\right) \int_0^R dr\,r \int_0^{2\pi}d\varphi\, e^{-i(l_2+l_3+1-l_1)\varphi} \left\{ J_{l_1}(p_{1\perp}r)J_{l_2}(p_{2\perp}r)J_{l_3+1}(p_{3\perp}r) \right.\nonumber \\
&\left. \times\left[ (\tilde{E}_{p_2}^{(\psi)}+m_f-p_{2z}-ip_{2\perp})(-\tilde{E}_{p_3}^{(\bar{\psi})}+m_f-p_{3z}-ip_{3\perp})+(-\tilde{E}_{p_2}^{(\psi)}+m_f+p_{2z}+ip_{2\perp})(\tilde{E}_{p_3}^{(\bar{\psi})}+m_f+p_{3z}+ip_{3\perp}) \right] \right. \nonumber \\
&\left. - \left[ (\tilde{E}_{p_2}^{(\psi)}+m_f+p_{2z}+ip_{2\perp})(-\tilde{E}_{p_3}^{(\bar{\psi})}+m_f+p_{3z}+ip_{3\perp})+(-\tilde{E}_{p_2}^{(\psi)}+m_f-p_{2z}-ip_{2\perp})(\tilde{E}_{p_3}^{(\bar{\psi})}+m_f-p_{3z}-ip_{3\perp})  \right] \right. \nonumber \\
&\left. \times J_{l_1}(p_{1\perp}r)J_{l_2+1}(p_{2\perp}r)J_{l_3}(p_{3\perp}r)\right\}.
\label{C1explicitbeforeintphi}
\end{align}
\end{widetext}

At this stage, the boundary conditions for the scalar and fermionic modes must be implemented separately. For the sigma field, we retain the Dirichlet boundary condition introduced above, $J_{l_1}(p_{1\perp}R)=0$, which discretizes its transverse momentum as $p_{1\perp}=\alpha_{l_1,n_1}/R$. For the fermionic modes, however, a Dirichlet condition cannot be imposed simultaneously on all spinor components, since the solutions contain both $J_l(p_\perp r)$ and $J_{l+1}(p_\perp r)$. Instead, we impose the spectral boundary conditions introduced in Ref.~\cite{Ambrus:2015lfr} for a Dirac field confined within a cylindrical boundary.

These boundary conditions follow from requiring the Dirac Hamiltonian to be self-adjoint. In particular, conservation of the inner product between arbitrary fermionic solutions requires the surface term associated with the radial Dirac current at $r=R$ to vanish, preventing fermionic probability flux through the cylindrical boundary. The same requirement is imposed on the charge-conjugated solutions. For the angular decomposition adopted here, these conditions imply that, for $l+\frac{1}{2}>0$, the first and third spinor components vanish at the boundary, whereas for $l+\frac{1}{2}<0$, the second and fourth components vanish. Consequently,
\begin{align}
J_l(p_\perp R)&=0, \qquad l+\frac{1}{2}>0, \nonumber \\
J_{l+1}(p_\perp R)&=0, \qquad l+\frac{1}{2}<0.
\label{fermionspectralBC}
\end{align}
The allowed transverse momenta can therefore be written as
\begin{equation}
p_{\perp,l,n}^{(f)}=\frac{\kappa_{l,n}}{R},
\label{fermiontransversemomentum}
\end{equation}
where
\begin{equation}
\kappa_{l,n}=
\begin{cases}
\alpha_{l,n}, & l+\frac{1}{2}>0,\\
\alpha_{-l-1,n}, & l+\frac{1}{2}<0,
\end{cases}
\label{fermionradialmodes}
\end{equation}
and $\alpha_{l,n}$ denotes the $n$th zero of the corresponding Bessel function. Notice that $\kappa_{l,n}$ fixes the transverse momentum of the entire fermionic mode and therefore appears as the argument of both $J_l$ and $J_{l+1}$ within the same spinor.

Accordingly, the vacuum dispersion relation for the discretized fermionic modes becomes
\begin{equation}
\tilde E_{nl}^{(f)}=\sqrt{p_z^2+m_f^2+\left(\frac{\kappa_{l,n}}{R}\right)^2}.
\label{fermiondiscretizedenergy}
\end{equation}

Finally, performing the angular integration and implementing the Dirichlet boundary condition for the sigma field together with the spectral boundary conditions for the fermionic modes, the vertex function becomes
\begin{widetext}
\begin{align}
V_{C1}&=-ig\, \delta\left( \tilde{E}_{n_1l_1}^{(\sigma)}-\tilde{E}_{n_2l_2}^{(\psi)}-\tilde{E}_{n_3l_3}^{(\bar\psi)}-\Omega(l_1-l_2-l_3-1)\right)
\delta\left(p_{1z}-p_{2z}-p_{3z}\right)\delta_{l_1,l_2+l_3+1} \nonumber \\
&\times\int_0^R dr\,r \left\{ J_{l_1}\left(\frac{\alpha_{l_1,n_1}}{R}r\right)J_{l_2}\left(\frac{\kappa_{l_2,n_2}}{R}r\right)J_{l_3+1}\left(\frac{\kappa_{l_3,n_3}}{R}r\right) \right.\nonumber \\
&\left. \times\left[
\left(\tilde{E}_{n_2l_2}^{(\psi)}+m_f-p_{2z}-i\frac{\kappa_{l_2,n_2}}{R}\right)
\left(-\tilde{E}_{n_3l_3}^{(\bar{\psi})}+m_f-p_{3z}-i\frac{\kappa_{l_3,n_3}}{R}\right)
\right.\right.\nonumber\\
&\left.\left. \qquad+
\left(-\tilde{E}_{n_2l_2}^{(\psi)}+m_f+p_{2z}+i\frac{\kappa_{l_2,n_2}}{R}\right)
\left(\tilde{E}_{n_3l_3}^{(\bar{\psi})}+m_f+p_{3z}+i\frac{\kappa_{l_3,n_3}}{R}\right)
\right] \right. \nonumber \\
&\left. - \left[
\left(\tilde{E}_{n_2l_2}^{(\psi)}+m_f+p_{2z}+i\frac{\kappa_{l_2,n_2}}{R}\right)
\left(-\tilde{E}_{n_3l_3}^{(\bar{\psi})}+m_f+p_{3z}+i\frac{\kappa_{l_3,n_3}}{R}\right)
\right.\right.\nonumber\\
&\left.\left. \qquad+
\left(-\tilde{E}_{n_2l_2}^{(\psi)}+m_f-p_{2z}-i\frac{\kappa_{l_2,n_2}}{R}\right)
\left(\tilde{E}_{n_3l_3}^{(\bar{\psi})}+m_f-p_{3z}-i\frac{\kappa_{l_3,n_3}}{R}\right)
\right] \right. \nonumber \\
&\left. \times J_{l_1}\left(\frac{\alpha_{l_1,n_1}}{R}r\right)J_{l_2+1}\left(\frac{\kappa_{l_2,n_2}}{R}r\right)J_{l_3}\left(\frac{\kappa_{l_3,n_3}}{R}r\right)\right\}.
\label{C1final}
\end{align}

We next consider the Yukawa interaction involving the neutral pion field, corresponding to diagram $C2$ in Fig.~\ref{fig3}. The derivation proceeds analogously to that of $C1$, with the sigma field replaced by the neutral pion field and the scalar coupling replaced by the corresponding pseudoscalar coupling involving $\gamma^5$. In this case, the fermionic contribution to the matrix element is proportional to $\bar u_{p_2}(r,\varphi)\gamma^5 v_{p_3}(r,\varphi)$. Evaluating this product explicitly using the spinors in Eqs.~(\ref{upspinor}) and~(\ref{vmspinor}), the $\gamma^5$ matrix is absorbed into the resulting spinor contraction, and the vertex function becomes
\begin{align}
    V_{C2}&=-g\, \delta\left( \tilde{E}_{n_1l_1}^{(\pi)}-\tilde{E}_{n_2l_2}^{(\psi)}-\tilde{E}_{n_3l_3}^{(\bar\psi)}-\Omega(l_1-l_2-l_3-1)\right)
\delta\left(p_{1z}-p_{2z}-p_{3z}\right)\delta_{l_1,l_2+l_3+1} \nonumber \\
&\times\int_0^R dr\,r \left\{ J_{l_1}\left(\frac{\alpha_{l_1,n_1}}{R}r\right)J_{l_2}\left(\frac{\kappa_{l_2,n_2}}{R}r\right)J_{l_3+1}\left(\frac{\kappa_{l_3,n_3}}{R}r\right) \right.\nonumber \\
&\left. \times\left[ \left(\tilde{E}_{n_2l_2}^{(\psi)}+m_f-p_{2z}-i\frac{\kappa_{l_2,n_2}}{R}\right)\left(\tilde{E}_{n_3l_3}^{(\bar{\psi})}+m_f+p_{3z}+i\frac{\kappa_{l_3,n_3}}{R}\right)\right. \right.\nonumber \\
&\left. \left. \qquad +\left(-\tilde{E}_{n_2l_2}^{(\psi)}+m_f+p_{2z}+i\frac{\kappa_{l_2,n_2}}{R}\right)\left(-\tilde{E}_{n_3l_3}^{(\bar{\psi})}+m_f-p_{3z}-i\frac{\kappa_{l_3,n_3}}{R}\right) \right] \right. \nonumber \\
&\left. - \left[ \left(\tilde{E}_{n_2l_2}^{(\psi)}+m_f+p_{2z}+i\frac{\kappa_{l_2,n_2}}{R}\right)\left(\tilde{E}_{n_3l_3}^{(\bar{\psi})}+m_f-p_{3z}-i\frac{\kappa_{l_3,n_3}}{R}\right)\right. \right.\nonumber \\
&\left. \left. \qquad +\left(-\tilde{E}_{n_2l_2}^{(\psi)}+m_f-p_{2z}-i\frac{\kappa_{l_2,n_2}}{R}\right)\left(-\tilde{E}_{n_3l_3}^{(\bar{\psi})}+m_f+p_{3z}+i\frac{\kappa_{l_3,n_3}}{R}\right)  \right] \right. \nonumber \\
&\left. \times J_{l_1}\left(\frac{\alpha_{l_1,n_1}}{R}r\right)J_{l_2+1}\left(\frac{\kappa_{l_2,n_2}}{R}r\right)J_{l_3}\left(\frac{\kappa_{l_3,n_3}}{R}r\right)\right\}.
\label{C2final}
\end{align}

Finally, we consider the Yukawa interaction involving the charged pion field, corresponding to diagram $C3$ in Fig.~\ref{fig3}. The derivation proceeds analogously to that of $C2$, with the only modification arising from the interaction vertex associated with the charged pion field. The resulting vertex function is
\begin{align}
    V_{C3}&=-\sqrt{2}g\, \delta\left( \tilde{E}_{n_1l_1}^{(\pi)}-\tilde{E}_{n_2l_2}^{(\psi)}-\tilde{E}_{n_3l_3}^{(\bar\psi)}-\Omega(l_1-l_2-l_3-1)\right)
\delta\left(p_{1z}-p_{2z}-p_{3z}\right)\delta_{l_1,l_2+l_3+1} \nonumber \\
&\times\int_0^R dr\,r \left\{ J_{l_1}\left(\frac{\alpha_{l_1,n_1}}{R}r\right)J_{l_2}\left(\frac{\kappa_{l_2,n_2}}{R}r\right)J_{l_3+1}\left(\frac{\kappa_{l_3,n_3}}{R}r\right) \right.\nonumber \\
&\left.  \times\left[ \left(\tilde{E}_{n_2l_2}^{(\psi)}+m_f-p_{2z}-i\frac{\kappa_{l_2,n_2}}{R}\right)\left(\tilde{E}_{n_3l_3}^{(\bar{\psi})}+m_f+p_{3z}+i\frac{\kappa_{l_3,n_3}}{R}\right)\right. \right.\nonumber \\
&\left. \left. \qquad +\left(-\tilde{E}_{n_2l_2}^{(\psi)}+m_f+p_{2z}+i\frac{\kappa_{l_2,n_2}}{R}\right)\left(-\tilde{E}_{n_3l_3}^{(\bar{\psi})}+m_f-p_{3z}-i\frac{\kappa_{l_3,n_3}}{R}\right) \right] \right. \nonumber \\
&\left. - \left[ \left(\tilde{E}_{n_2l_2}^{(\psi)}+m_f+p_{2z}+i\frac{\kappa_{l_2,n_2}}{R}\right)\left(\tilde{E}_{n_3l_3}^{(\bar{\psi})}+m_f-p_{3z}-i\frac{\kappa_{l_3,n_3}}{R}\right)\right. \right.\nonumber \\
&\left. \left. \qquad +\left(-\tilde{E}_{n_2l_2}^{(\psi)}+m_f-p_{2z}-i\frac{\kappa_{l_2,n_2}}{R}\right)\left(-\tilde{E}_{n_3l_3}^{(\bar{\psi})}+m_f+p_{3z}+i\frac{\kappa_{l_3,n_3}}{R}\right)  \right] \right. \nonumber \\
&\left. \times J_{l_1}\left(\frac{\alpha_{l_1,n_1}}{R}r\right)J_{l_2+1}\left(\frac{\kappa_{l_2,n_2}}{R}r\right)J_{l_3}\left(\frac{\kappa_{l_3,n_3}}{R}r\right)\right\}.
\label{C3final}
\end{align}
\end{widetext}

The results obtained in this section show that global rotation modifies the momentum-space structure of all interaction vertices of the LSMq. In contrast to the usual vacuum Feynman rules, the vertices can no longer be represented by a four-dimensional Dirac delta enforcing conservation of the four-momentum. Instead, the symmetries preserved by the rotating background lead to the conservation of the longitudinal momentum and the total angular momentum along the rotation axis, while the transverse dynamics is encoded in nontrivial radial form factors involving products of Bessel functions. The causality restriction, together with the boundary conditions at $r=R$, further discretizes the transverse momenta. For the scalar fields, this discretization follows from the Dirichlet boundary condition, whereas for the fermionic modes it results from the spectral boundary conditions required for a consistent formulation of the Dirac field within the cylindrical boundary. For the Yukawa vertices, the angular-momentum selection rule explicitly incorporates the fermionic spin contribution, leading to $l_1=l_2+l_3+1$ for the processes considered above. An important feature of these results is the direct connection between the rotational shift of the energies and the conservation of angular momentum. The same combination of angular-momentum quantum numbers that multiplies $\Omega$ in the energy-conservation condition is constrained to vanish by the corresponding angular-momentum selection rule. Thus, the effects of rotation remain encoded in the structure of the allowed modes, their boundary-induced discrete transverse spectra, and the resulting radial form factors.


\section{\label{sec5} Conclusions}

In this work, we have derived the interaction vertices of the linear sigma model with quarks in the presence of global rotation. Starting from the solutions of the scalar and Dirac equations in a rotating background, we used their mode expansions to evaluate the interaction matrix elements associated with the mesonic self-interactions and the Yukawa couplings of the model. In this way, the modifications induced by rotation emerge directly from the structure of the field modes, rather than being introduced through an effective prescription for the interaction vertices.

A common structure emerges from the different classes of vertices considered. Unlike in vacuum, where translational invariance leads to the usual four-dimensional Dirac delta enforcing four-momentum conservation, global rotation singles out a preferred axis and changes the transverse structure of the interaction. Energy and longitudinal momentum remain conserved, while rotational symmetry around the $z$ axis leads to angular-momentum selection rules. The transverse dynamics is instead contained in radial form factors involving products of Bessel functions. The finite radial boundary further discretizes the transverse spectrum, with the allowed modes determined by the corresponding boundary conditions. For the scalar fields, Dirichlet boundary conditions lead to a transverse spectrum determined by the zeros of the appropriate Bessel functions, whereas for the fermionic fields we impose spectral boundary conditions consistent with the self-adjointness of the Dirac Hamiltonian and charge conjugation. For the Yukawa interactions, the resulting angular-momentum selection rules explicitly incorporate the spin contribution of the fermionic modes. The resulting structure provides a consistent set of momentum-space interaction rules adapted to the symmetries and finite geometry of the rotating system.

Although the derivation has been carried out within the LSMq, the results have a broader scope. The structures obtained for the purely mesonic vertices are relevant to theories containing self-interacting scalar fields, whereas the fermion-scalar vertices provide the corresponding building blocks for theories with Yukawa-type interactions in a rotating background. Therefore, the expressions derived here are not restricted to the particular phenomenological content of the LSMq, but provide a framework that can be adapted to a wider class of quantum field theories involving scalar and fermionic degrees of freedom under global rotation.

These results provide the ingredients required to extend the analysis of rotating quantum systems beyond the level of free propagators and thermodynamic potentials. In particular, the interaction vertices derived here can be used as building blocks for perturbative calculations involving loop corrections, self-energies, scattering amplitudes, and in-medium processes in rotating backgrounds. Such developments are necessary for constructing a more complete microscopic description, within quantum field theory, of strongly interacting matter subject to global rotation, as encountered in relativistic heavy-ion collisions and other rotating many-body systems.

\begin{acknowledgments}
Support for this work was received by the Secretaria de Ciencia, Humanidades, Tecnología e Innovación Grant No. CBF-2025-G-1718. R. Z. acknowledges support from the following grant ANID/FONDECYT/Regular No. 1241436. E. L. acknowledges the financial support of a fellowship granted by Secretaria de Ciencia, Humanidades, Tecnología e Innovación as part of the Sistema Nacional de Posgrados.
\end{acknowledgments}

\bibliography{mybibliography}

@article{Ambrus:2014uqa,
    author = "Ambru\c{s}, Victor E. and Winstanley, Elizabeth",
    title = "{Rotating quantum states}",
    eprint = "1401.6388",
    archivePrefix = "arXiv",
    primaryClass = "hep-th",
    doi = "10.1016/j.physletb.2014.05.031",
    journal = "Phys. Lett. B",
    volume = "734",
    pages = "296--301",
    year = "2014"
}

@article{Ayala:2021osy,
    author = "Ayala, Alejandro and Hern{\'a}ndez, L. A. and Raya, K. and Zamora, R.",
    title = "{Fermion propagator in a rotating environment}",
    eprint = "2102.03476",
    archivePrefix = "arXiv",
    primaryClass = "hep-ph",
    doi = "10.1103/PhysRevD.104.039901",
    journal = "Phys. Rev. D",
    volume = "103",
    number = "7",
    pages = "076021",
    year = "2021",
    note = "[Erratum: Phys.Rev.D 104, 039901 (2021)]"
}

@article{STAR:2017ckg,
    author = "Adamczyk, L. and others",
    collaboration = "STAR",
    title = "{Global $\Lambda$ hyperon polarization in nuclear collisions: evidence for the most vortical fluid}",
    eprint = "1701.06657",
    archivePrefix = "arXiv",
    primaryClass = "nucl-ex",
    doi = "10.1038/nature23004",
    journal = "Nature",
    volume = "548",
    pages = "62--65",
    year = "2017"
}

@article{Deng:2020ygd,
    author = "Deng, Xian-Gai and Huang, Xu-Guang and Ma, Yu-Gang and Zhang, Song",
    title = "{Vorticity in low-energy heavy-ion collisions}",
    eprint = "2001.01371",
    archivePrefix = "arXiv",
    primaryClass = "nucl-th",
    doi = "10.1103/PhysRevC.101.064908",
    journal = "Phys. Rev. C",
    volume = "101",
    number = "6",
    pages = "064908",
    year = "2020"
}

@article{Deng:2016gyh,
    author = "Deng, Wei-Tian and Huang, Xu-Guang",
    title = "{Vorticity in Heavy-Ion Collisions}",
    eprint = "1603.06117",
    archivePrefix = "arXiv",
    primaryClass = "nucl-th",
    doi = "10.1103/PhysRevC.93.064907",
    journal = "Phys. Rev. C",
    volume = "93",
    number = "6",
    pages = "064907",
    year = "2016"
}

@unpublished{Sahoo:2026lrw,
    author = "Sahoo, Bhagyarathi and Singh, Captain R. and Sahoo, Raghunath",
    title = "{Probing Rotational Dynamics of Quark Gluon Plasma via Global Vorticity}",
    eprint = "2602.13618",
    archivePrefix = "arXiv",
    primaryClass = "hep-ph",
    month = "2",
    year = "2026"
}

@unpublished{Savchuk:2025kuk,
    author = "Savchuk, Oleh and Danielewicz, Pawel and Kincses, Daniel and Sorensen, Agnieszka",
    title = "{Probing vorticity through femtoscopic correlations}",
    eprint = "2510.10795",
    archivePrefix = "arXiv",
    primaryClass = "nucl-th",
    month = "10",
    year = "2025"
}

@unpublished{Ayala:2025kuc,
    author = "Ayala, Alejandro and Medina-Serna, Jos{\'e} Jorge and Dom{\'\i}nguez, Isabel and Maldonado, Ivonne and Tejeda-Yeomans, Mar{\'\i}a Elena",
    title = "{Thermal and baryon density modifications to the $\sigma$-boson propagator: A road to describe the transfer of vorticity to spin in a nuclear environment in relativistic heavy-ion collisions}",
    eprint = "2502.14248",
    archivePrefix = "arXiv",
    primaryClass = "nucl-th",
    month = "2",
    year = "2025"
}

@article{Huang:2024aob,
    author = "Huang, Anping and Wu, Xiang-Yu and Huang, Mei",
    title = "{Effect of vorticity on the dynamical magnetic fields in heavy-ion collisions}",
    eprint = "2408.14077",
    archivePrefix = "arXiv",
    primaryClass = "hep-ph",
    doi = "10.1103/PhysRevD.110.094032",
    journal = "Phys. Rev. D",
    volume = "110",
    number = "9",
    pages = "094032",
    year = "2024"
}

@article{Gao:2023wwo,
    author = "Gao, Jian-Hua and Yang, Shi-Zheng",
    title = "{Revisiting the spin effects induced by thermal vorticity}",
    eprint = "2308.16616",
    archivePrefix = "arXiv",
    primaryClass = "hep-ph",
    doi = "10.1088/1674-1137/ad2363",
    journal = "Chin. Phys. C",
    volume = "48",
    number = "5",
    pages = "053114",
    year = "2024"
}

@article{Sahoo:2023xnu,
    author = "Sahoo, Bhagyarathi and Singh, Captain R. and Sahu, Dushmanta and Sahoo, Raghunath and Alam, Jan-e",
    title = "{Impact of vorticity and viscosity on the hydrodynamic evolution of hot QCD medium}",
    eprint = "2302.07668",
    archivePrefix = "arXiv",
    primaryClass = "hep-ph",
    doi = "10.1140/epjc/s10052-023-12027-3",
    journal = "Eur. Phys. J. C",
    volume = "83",
    number = "9",
    pages = "873",
    year = "2023"
}

@article{Tsegelnik:2022eoz,
    author = "Tsegelnik, N. S. and Kolomeitsev, E. E. and Voronyuk, V.",
    title = "{Helicity and vorticity in heavy-ion collisions at energies available at the JINR Nuclotron-based Ion Collider facility}",
    eprint = "2211.09219",
    archivePrefix = "arXiv",
    primaryClass = "nucl-th",
    doi = "10.1103/PhysRevC.107.034906",
    journal = "Phys. Rev. C",
    volume = "107",
    number = "3",
    pages = "034906",
    year = "2023"
}

@article{Huang:2020dtn,
    author = "Huang, Xu-Guang and Liao, Jinfeng and Wang, Qun and Xia, Xiao-Liang",
    title = "{Vorticity and Spin Polarization in Heavy Ion Collisions: Transport Models}",
    eprint = "2010.08937",
    archivePrefix = "arXiv",
    primaryClass = "nucl-th",
    doi = "10.1007/978-3-030-71427-7_9",
    journal = "Lect. Notes Phys.",
    volume = "987",
    pages = "281--308",
    year = "2021"
}

@article{Ayala:2020ndx,
    author = "Ayala, Alejandro and de la Cruz, David and Hern{\'a}ndez, L. A. and Salinas, Jordi",
    title = "{Relaxation time for the alignment between the spin of a finite-mass quark or antiquark and the thermal vorticity in relativistic heavy-ion collisions}",
    eprint = "2003.06545",
    archivePrefix = "arXiv",
    primaryClass = "hep-ph",
    doi = "10.1103/PhysRevD.102.056019",
    journal = "Phys. Rev. D",
    volume = "102",
    number = "5",
    pages = "056019",
    year = "2020"
}

@article{Xie:2015xpa,
    author = "Xie, Yilong and Glastad, Robert C. and Csernai, L\'aszl\'o P.",
    title = "{$\Lambda$ polarization in an exact rotating and expanding fluid dynamical model for peripheral heavy ion reactions}",
    eprint = "1505.07221",
    archivePrefix = "arXiv",
    primaryClass = "nucl-th",
    doi = "10.1103/PhysRevC.92.064901",
    journal = "Phys. Rev. C",
    volume = "92",
    number = "6",
    pages = "064901",
    year = "2015"
}

@article{Xie:2016fjj,
    author = {Xie, Y. L. and Bleicher, M. and St\"ocker, H. and Wang, D. J. and Csernai, L. P.},
    title = "{$\Lambda$ polarization in peripheral collisions at moderate relativistic energies}",
    eprint = "1610.08678",
    archivePrefix = "arXiv",
    primaryClass = "nucl-th",
    doi = "10.1103/PhysRevC.94.054907",
    journal = "Phys. Rev. C",
    volume = "94",
    number = "5",
    pages = "054907",
    year = "2016"
}

@article{Becattini:2016gvu,
    author = "Becattini, F. and Karpenko, I. and Lisa, M. and Upsal, I. and Voloshin, S.",
    title = "{Global hyperon polarization at local thermodynamic equilibrium with vorticity, magnetic field and feed-down}",
    eprint = "1610.02506",
    archivePrefix = "arXiv",
    primaryClass = "nucl-th",
    doi = "10.1103/PhysRevC.95.054902",
    journal = "Phys. Rev. C",
    volume = "95",
    number = "5",
    pages = "054902",
    year = "2017"
}

@article{Sorin:2016smp,
    author = "Sorin, Alexander and Teryaev, Oleg",
    title = "{Axial anomaly and energy dependence of hyperon polarization in Heavy-Ion Collisions}",
    eprint = "1606.08398",
    archivePrefix = "arXiv",
    primaryClass = "nucl-th",
    doi = "10.1103/PhysRevC.95.011902",
    journal = "Phys. Rev. C",
    volume = "95",
    number = "1",
    pages = "011902",
    year = "2017"
}

@article{Li:2017slc,
    author = "Li, Hui and Pang, Long-Gang and Wang, Qun and Xia, Xiao-Liang",
    title = "{Global $\Lambda$ polarization in heavy-ion collisions from a transport model}",
    eprint = "1704.01507",
    archivePrefix = "arXiv",
    primaryClass = "nucl-th",
    doi = "10.1103/PhysRevC.96.054908",
    journal = "Phys. Rev. C",
    volume = "96",
    number = "5",
    pages = "054908",
    year = "2017"
}

@article{Sun:2017xhx,
    author = "Sun, Yifeng and Ko, Che Ming",
    title = "{$\Lambda$ hyperon polarization in relativistic heavy ion collisions from a chiral kinetic approach}",
    eprint = "1706.09467",
    archivePrefix = "arXiv",
    primaryClass = "nucl-th",
    doi = "10.1103/PhysRevC.96.024906",
    journal = "Phys. Rev. C",
    volume = "96",
    number = "2",
    pages = "024906",
    year = "2017"
}

@article{Han:2017hdi,
    author = "Han, Zhang-Zhu and Xu, Jun",
    title = "{Investigating different $\Lambda$ and $\bar\Lambda$ polarizations in relativistic heavy-ion collisions}",
    eprint = "1707.07262",
    archivePrefix = "arXiv",
    primaryClass = "nucl-th",
    doi = "10.1016/j.physletb.2018.10.001",
    journal = "Phys. Lett. B",
    volume = "786",
    pages = "255--259",
    year = "2018"
}

@article{Xia:2018tes,
    author = "Xia, Xiao-Liang and Li, Hui and Tang, Ze-Bo and Wang, Qun",
    title = "{Probing vorticity structure in heavy-ion collisions by local $\Lambda$ polarization}",
    eprint = "1803.00867",
    archivePrefix = "arXiv",
    primaryClass = "nucl-th",
    doi = "10.1103/PhysRevC.98.024905",
    journal = "Phys. Rev. C",
    volume = "98",
    pages = "024905",
    year = "2018"
}

@article{Baznat:2017jfj,
    author = "Baznat, Mircea and Gudima, Konstantin and Sorin, Alexander and Teryaev, Oleg",
    title = "{Hyperon polarization in heavy-ion collisions and holographic gravitational anomaly}",
    eprint = "1701.00923",
    archivePrefix = "arXiv",
    primaryClass = "nucl-th",
    doi = "10.1103/PhysRevC.97.041902",
    journal = "Phys. Rev. C",
    volume = "97",
    number = "4",
    pages = "041902",
    year = "2018"
}

@article{Karpenko:2018erl,
    author = "Karpenko, Iurii and Becattini, Francesco",
    editor = "Antinori, Federico and Dainese, Andrea and Giubellino, Paolo and Greco, Vincenzo and Lombardo, Maria Paola and Scomparin, Enrico",
    title = "{Lambda polarization in heavy ion collisions: from RHIC BES to LHC energies}",
    eprint = "1811.00322",
    archivePrefix = "arXiv",
    primaryClass = "nucl-th",
    doi = "10.1016/j.nuclphysa.2018.10.067",
    journal = "Nucl. Phys. A",
    volume = "982",
    pages = "519--522",
    year = "2019"
}

@article{Suvarieva:2018wmh,
    author = "Suvarieva, Dilyana and Gudima, K. and Zinchenko, A.",
    title = "{A Monte Carlo Study of Lambda Hyperon Polarization at BM@N}",
    doi = "10.1134/S1547477118020115",
    journal = "Phys. Part. Nucl. Lett.",
    volume = "15",
    number = "2",
    pages = "182--188",
    year = "2018"
}

@article{Kolomeitsev:2018svb,
    author = "Kolomeitsev, E. E. and Toneev, V. D. and Voronyuk, V.",
    title = "{Vorticity and hyperon polarization at energies available at JINR Nuclotron-based Ion Collider fAcility}",
    eprint = "1801.07610",
    archivePrefix = "arXiv",
    primaryClass = "nucl-th",
    doi = "10.1103/PhysRevC.97.064902",
    journal = "Phys. Rev. C",
    volume = "97",
    number = "6",
    pages = "064902",
    year = "2018"
}

@article{Xie:2019jun,
    author = "Xie, Yilong and Wang, Dujuan and Csernai, Laszlo Pal",
    title = "{Fluid dynamics study of the $\varLambda $ polarization for Au + Au collisions at $\sqrt{s_{NN}}=200$ GeV}",
    eprint = "1907.00773",
    archivePrefix = "arXiv",
    primaryClass = "hep-ph",
    doi = "10.1140/epjc/s10052-019-7576-8",
    journal = "Eur. Phys. J. C",
    volume = "80",
    number = "1",
    pages = "39",
    year = "2020"
}

@article{Ayala:2021xrn,
    author = "Ayala, Alejandro and Dom\'\i{}nguez, Isabel and Maldonado, Ivonne and Tejeda-Yeomans, Mar\'\i{}a Elena",
    title = "{Rise and fall of \ensuremath{\Lambda} and \ensuremath{\Lambda}\textasciimacron{} global polarization in semi-central heavy-ion collisions at HADES, NICA and RHIC energies from the core-corona model}",
    eprint = "2106.14379",
    archivePrefix = "arXiv",
    primaryClass = "hep-ph",
    doi = "10.1103/PhysRevC.105.034907",
    journal = "Phys. Rev. C",
    volume = "105",
    number = "3",
    pages = "034907",
    year = "2022"
}

@article{Ayala:2020soy,
    author = "Ayala, Alejandro and others",
    title = "{Core meets corona: A two-component source to explain $\Lambda$ and $\bar \Lambda$ global polarization in semi-central heavy-ion collisions}",
    eprint = "2003.13757",
    archivePrefix = "arXiv",
    primaryClass = "hep-ph",
    doi = "10.1016/j.physletb.2020.135818",
    journal = "Phys. Lett. B",
    volume = "810",
    pages = "135818",
    year = "2020"
}

@unpublished{Padhan:2026mwg,
    author = "Padhan, Nandita and Pradhan, Kshitish Kumar and Chatterjee, Arghya and Sahoo, Raghunath",
    title = "{Vorticity-induced modifications of chemical freeze-out in heavy-ion collisions}",
    eprint = "2603.27267",
    archivePrefix = "arXiv",
    primaryClass = "hep-ph",
    month = "3",
    year = "2026"
}

@article{Troshin:2025fko,
    author = "Troshin, V. V.",
    title = "{Global Polarization and Collective Flow of $\Lambda$ Hyperons in the MPD Experiment}",
    doi = "10.1134/S1063779624702496",
    journal = "Phys. Part. Nucl.",
    volume = "56",
    number = "3",
    pages = "903--908",
    year = "2025"
}

@article{STAR:2025dgs,
    author = "Aboona, B. E. and others",
    collaboration = "STAR",
    title = "{Hyperon global polarization in isobar Ru+Ru and Zr+Zr collisions at sNN=200GeV}",
    eprint = "2505.05046",
    archivePrefix = "arXiv",
    primaryClass = "nucl-ex",
    doi = "10.1016/j.physletb.2025.139891",
    journal = "Phys. Lett. B",
    volume = "870",
    pages = "139891",
    year = "2025"
}

@article{Ivanov:2025izv,
    author = "Ivanov, Yu. B.",
    title = "{Global {\ensuremath{\Lambda}} polarization in heavy-ion collisions at high baryon density}",
    eprint = "2504.12200",
    archivePrefix = "arXiv",
    primaryClass = "nucl-th",
    doi = "10.1103/qn56-fvnq",
    journal = "Phys. Rev. C",
    volume = "112",
    number = "1",
    pages = "014902",
    year = "2025"
}

@article{Sapna:2025yss,
    author = "Sapna and Singh, Sushant K. and Wagner, David",
    title = "{Spin polarization of {\ensuremath{\Lambda}} hyperons from dissipative spin hydrodynamics}",
    eprint = "2503.22552",
    archivePrefix = "arXiv",
    primaryClass = "hep-ph",
    doi = "10.1103/1s6g-fs8w",
    journal = "Phys. Rev. C",
    volume = "112",
    number = "5",
    pages = "054902",
    year = "2025"
}

@article{Palermo:2024tza,
    author = "Palermo, Andrea and Grossi, Eduardo and Karpenko, Iurii and Becattini, Francesco",
    title = "{$\Lambda $ polarization in very high energy heavy ion collisions as a probe of the quark{\textendash}gluon plasma formation and properties}",
    eprint = "2404.14295",
    archivePrefix = "arXiv",
    primaryClass = "nucl-th",
    doi = "10.1140/epjc/s10052-024-13229-z",
    journal = "Eur. Phys. J. C",
    volume = "84",
    number = "9",
    pages = "920",
    year = "2024"
}

@article{STAR:2023nvo,
    author = "Abdulhamid, M. I. and others",
    collaboration = "STAR",
    title = "{Global polarization of {\ensuremath{\Lambda}} and {\ensuremath{\Lambda}}{\textasciimacron} hyperons in Au+Au collisions at sNN=19.6 and 27 GeV}",
    eprint = "2305.08705",
    archivePrefix = "arXiv",
    primaryClass = "nucl-ex",
    doi = "10.1103/PhysRevC.108.014910",
    journal = "Phys. Rev. C",
    volume = "108",
    number = "1",
    pages = "014910",
    year = "2023"
}

@article{Ayala:2023xyn,
    author = "Ayala, Alejandro and Dominguez, Isabel and Maldonado, Ivonne and Tejeda-Yeomans, Maria Elena",
    title = "{An Improved Core-Corona Model for {\ensuremath{\Lambda}} and {\ensuremath{\Lambda}} Polarization in Relativistic Heavy-Ion Collisions}",
    doi = "10.3390/particles6010022",
    journal = "Particles",
    volume = "6",
    number = "1",
    pages = "405--415",
    year = "2023"
}

@article{Hernandez:2024nev,
    author = "Hern{\'a}ndez, Luis A. and Zamora, R.",
    title = "{Vortical effects and the critical end point in the linear sigma model coupled to quark}",
    eprint = "2410.17874",
    archivePrefix = "arXiv",
    primaryClass = "hep-ph",
    doi = "10.1103/PhysRevD.111.036003",
    journal = "Phys. Rev. D",
    volume = "111",
    number = "3",
    pages = "036003",
    year = "2025"
}

@unpublished{Siri:2025kdw,
    author = "Siri, E. and Sadooghi, N.",
    title = "{Spontaneous breaking of global U(1) symmetry in an interacting Bose gas under rigid rotation}",
    eprint = "2508.17055",
    archivePrefix = "arXiv",
    primaryClass = "hep-ph",
    month = "8",
    year = "2025"
}

@article{Farias:2025vss,
    author = "Farias, Ricardo L. S. and Tavares, William R.",
    title = "{Chiral and deconfinement transitions in spin-polarized quark matter}",
    eprint = "2507.08130",
    archivePrefix = "arXiv",
    primaryClass = "hep-ph",
    doi = "10.1103/5znc-7ztg",
    journal = "Phys. Rev. D",
    volume = "112",
    number = "5",
    pages = "L051902",
    year = "2025"
}

@article{Zhu:2025pxh,
    author = "Zhu, Zhi-Bin and Chen, Hao-Lei and Huang, Xu-Guang",
    title = "{Chiral symmetry breaking in accelerating and rotating frames}",
    eprint = "2511.03230",
    archivePrefix = "arXiv",
    primaryClass = "hep-ph",
    doi = "10.1103/yxj9-33z6",
    journal = "Phys. Rev. D",
    volume = "113",
    number = "3",
    pages = "034005",
    year = "2026"
}

@article{Gaspar:2023nqk,
    author = "Gaspar, Irving I. and Hern{\'a}ndez, Luis A. and Zamora, Renato",
    title = "{Chiral symmetry restoration in a rotating medium}",
    eprint = "2305.00101",
    archivePrefix = "arXiv",
    primaryClass = "hep-ph",
    doi = "10.1103/PhysRevD.108.094020",
    journal = "Phys. Rev. D",
    volume = "108",
    number = "9",
    pages = "094020",
    year = "2023"
}

@article{Jiang:2016wvv,
    author = "Jiang, Yin and Liao, Jinfeng",
    title = "{Pairing Phase Transitions of Matter under Rotation}",
    eprint = "1606.03808",
    archivePrefix = "arXiv",
    primaryClass = "hep-ph",
    doi = "10.1103/PhysRevLett.117.192302",
    journal = "Phys. Rev. Lett.",
    volume = "117",
    number = "19",
    pages = "192302",
    year = "2016"
}

@article{Wang:2018sur,
    author = "Wang, Xinyang and Wei, Minghua and Li, Zhibin and Huang, Mei",
    title = "{Quark matter under rotation in the NJL model with vector interaction}",
    eprint = "1808.01931",
    archivePrefix = "arXiv",
    primaryClass = "hep-ph",
    doi = "10.1103/PhysRevD.99.016018",
    journal = "Phys. Rev. D",
    volume = "99",
    number = "1",
    pages = "016018",
    year = "2019"
}

@article{Chernodub:2016kxh,
    author = "Chernodub, M. N. and Gongyo, Shinya",
    title = "{Interacting fermions in rotation: chiral symmetry restoration, moment of inertia and thermodynamics}",
    eprint = "1611.02598",
    archivePrefix = "arXiv",
    primaryClass = "hep-th",
    reportNumber = "RIKEN-QHP-255",
    doi = "10.1007/JHEP01(2017)136",
    journal = "JHEP",
    volume = "01",
    pages = "136",
    year = "2017"
}

@article{Chen:2023cjt,
    author = "Chen, Hao-Lei and Zhu, Zhi-Bin and Huang, Xu-Guang",
    title = "{Quark-meson model under rotation: A functional renormalization group study}",
    eprint = "2306.08362",
    archivePrefix = "arXiv",
    primaryClass = "hep-ph",
    doi = "10.1103/PhysRevD.108.054006",
    journal = "Phys. Rev. D",
    volume = "108",
    number = "5",
    pages = "054006",
    year = "2023"
}

@article{Sun:2023kuu,
    author = "Sun, Fei and Xu, Kun and Huang, Mei",
    title = "{Splitting of chiral and deconfinement phase transitions induced by rotation}",
    eprint = "2307.14402",
    archivePrefix = "arXiv",
    primaryClass = "hep-ph",
    doi = "10.1103/PhysRevD.108.096007",
    journal = "Phys. Rev. D",
    volume = "108",
    number = "9",
    pages = "096007",
    year = "2023"
}

@article{Fujimoto:2021xix,
    author = "Fujimoto, Yuki and Fukushima, Kenji and Hidaka, Yoshimasa",
    title = "{Deconfining Phase Boundary of Rapidly Rotating Hot and Dense Matter and Analysis of Moment of Inertia}",
    eprint = "2101.09173",
    archivePrefix = "arXiv",
    primaryClass = "hep-ph",
    reportNumber = "KEK-TH-2290, J-PARC-TH-0236, RIKEN-iTHEMS-Report-21",
    doi = "10.1016/j.physletb.2021.136184",
    journal = "Phys. Lett. B",
    volume = "816",
    pages = "136184",
    year = "2021"
}

@article{Chen:2020ath,
    author = "Chen, Xun and Zhang, Lin and Li, Danning and Hou, Defu and Huang, Mei",
    title = "{Gluodynamics and deconfinement phase transition under rotation from holography}",
    eprint = "2010.14478",
    archivePrefix = "arXiv",
    primaryClass = "hep-ph",
    doi = "10.1007/JHEP07(2021)132",
    journal = "JHEP",
    volume = "07",
    pages = "132",
    year = "2021"
}

@article{Braga:2022yfe,
    author = "Braga, Nelson R. F. and Faulhaber, Luiz F. and Junqueira, Octavio C.",
    title = "{Confinement-deconfinement temperature for a rotating quark-gluon plasma}",
    eprint = "2201.05581",
    archivePrefix = "arXiv",
    primaryClass = "hep-th",
    doi = "10.1103/PhysRevD.105.106003",
    journal = "Phys. Rev. D",
    volume = "105",
    number = "10",
    pages = "106003",
    year = "2022"
}

@article{Wang:2024szr,
    author = "Wang, Jia-Hao and Feng, Sheng-Qin",
    title = "{Rotation effect on the deconfinement phase transition in holographic QCD}",
    eprint = "2403.01814",
    archivePrefix = "arXiv",
    primaryClass = "hep-ph",
    doi = "10.1103/PhysRevD.109.066019",
    journal = "Phys. Rev. D",
    volume = "109",
    number = "6",
    pages = "066019",
    year = "2024"
}

@article{Braguta:2020biu,
    author = "Braguta, V. V. and Kotov, A. Yu. and Kuznedelev, D. D. and Roenko, A. A.",
    title = "{Study of the Confinement/Deconfinement Phase Transition in Rotating Lattice SU(3) Gluodynamics}",
    doi = "10.31857/S1234567820130029",
    journal = "Pisma Zh. Eksp. Teor. Fiz.",
    volume = "112",
    number = "1",
    pages = "9--16",
    year = "2020"
}

@article{Braguta:2021jgn,
    author = "Braguta, V. V. and Kotov, A. Yu. and Kuznedelev, D. D. and Roenko, A. A.",
    title = "{Influence of relativistic rotation on the confinement-deconfinement transition in gluodynamics}",
    eprint = "2102.05084",
    archivePrefix = "arXiv",
    primaryClass = "hep-lat",
    doi = "10.1103/PhysRevD.103.094515",
    journal = "Phys. Rev. D",
    volume = "103",
    number = "9",
    pages = "094515",
    year = "2021"
}

@unpublished{Yang:2023vsw,
    author = "Yang, Ji-Chong and Huang, Xu-Guang",
    title = "{QCD on Rotating Lattice with Staggered Fermions}",
    eprint = "2307.05755",
    archivePrefix = "arXiv",
    primaryClass = "hep-lat",
    month = "7",
    year = "2023"
}

@article{Sun:2024anu,
    author = "Sun, Fei and Shao, Jingdong and Wen, Rui and Xu, Kun and Huang, Mei",
    title = "{Chiral phase transition and spin alignment of vector mesons in the polarized-Polyakov-loop Nambu{\textendash}Jona-Lasinio model under rotation}",
    eprint = "2402.16595",
    archivePrefix = "arXiv",
    primaryClass = "hep-ph",
    doi = "10.1103/PhysRevD.109.116017",
    journal = "Phys. Rev. D",
    volume = "109",
    number = "11",
    pages = "116017",
    year = "2024"
}

@article{Cao:2023olg,
    author = "Cao, Gaoqing",
    title = "{Effects of imaginary and real rotations on QCD matters}",
    eprint = "2310.03310",
    archivePrefix = "arXiv",
    primaryClass = "nucl-th",
    doi = "10.1103/PhysRevD.109.014001",
    journal = "Phys. Rev. D",
    volume = "109",
    number = "1",
    pages = "014001",
    year = "2024"
}

@article{Jiang:2023zzu,
    author = "Jiang, Yin",
    title = "{Rotating SU(2) gluon matter and deconfinement at finite temperature}",
    eprint = "2312.06166",
    archivePrefix = "arXiv",
    primaryClass = "hep-th",
    doi = "10.1016/j.physletb.2024.138655",
    journal = "Phys. Lett. B",
    volume = "853",
    pages = "138655",
    year = "2024"
}

@article{TabatabaeeMehr:2023tpt,
    author = "Tabatabaee Mehr, S. M. A.",
    title = "{Chiral symmetry breaking and phase diagram of dual chiral density wave in a rotating quark matter}",
    eprint = "2306.11753",
    archivePrefix = "arXiv",
    primaryClass = "nucl-th",
    doi = "10.1103/PhysRevD.108.094042",
    journal = "Phys. Rev. D",
    volume = "108",
    number = "9",
    pages = "094042",
    year = "2023"
}

@article{Chernodub:2020qah,
    author = "Chernodub, M. N.",
    title = "{Inhomogeneous confining-deconfining phases in rotating plasmas}",
    eprint = "2012.04924",
    archivePrefix = "arXiv",
    primaryClass = "hep-ph",
    doi = "10.1103/PhysRevD.103.054027",
    journal = "Phys. Rev. D",
    volume = "103",
    number = "5",
    pages = "054027",
    year = "2021"
}

@article{Chernodub:2022veq,
    author = "Chernodub, M. N. and Goy, V. A. and Molochkov, A. V.",
    title = "{Inhomogeneity of a rotating gluon plasma and the Tolman-Ehrenfest law in imaginary time: Lattice results for fast imaginary rotation}",
    eprint = "2209.15534",
    archivePrefix = "arXiv",
    primaryClass = "hep-lat",
    doi = "10.1103/PhysRevD.107.114502",
    journal = "Phys. Rev. D",
    volume = "107",
    number = "11",
    pages = "114502",
    year = "2023"
}

@article{Braguta:2023iyx,
    author = "Braguta, Victor V. and Chernodub, Maxim N. and Roenko, Artem A.",
    title = "{New mixed inhomogeneous phase in vortical gluon plasma: First-principle results from rotating SU(3) lattice gauge theory}",
    eprint = "2312.13994",
    archivePrefix = "arXiv",
    primaryClass = "hep-lat",
    doi = "10.1016/j.physletb.2024.138783",
    journal = "Phys. Lett. B",
    volume = "855",
    pages = "138783",
    year = "2024"
}

@article{Palermo:2023cup,
    author = "Palermo, Andrea and Becattini, Francesco",
    title = "{Exact spin polarization of massive and massless particles in relativistic fluids at global equilibrium}",
    eprint = "2304.02276",
    archivePrefix = "arXiv",
    primaryClass = "nucl-th",
    doi = "10.1140/epjp/s13360-023-04169-w",
    journal = "Eur. Phys. J. Plus",
    volume = "138",
    number = "6",
    pages = "547",
    year = "2023"
}

@article{Xie:2023gbo,
    author = "Xie, Xin-Qing and Wang, Dong-Lin and Yang, Chen and Pu, Shi",
    title = "{Causality and stability analysis for the minimal causal spin hydrodynamics}",
    eprint = "2306.13880",
    archivePrefix = "arXiv",
    primaryClass = "hep-ph",
    doi = "10.1103/PhysRevD.108.094031",
    journal = "Phys. Rev. D",
    volume = "108",
    number = "9",
    pages = "094031",
    year = "2023"
}

@article{Fang:2025aig,
    author = "Fang, Shuo and Fukushima, Kenji and Pu, Shi and Wang, Dong-Lin",
    title = "{Relativistic Spin Hydrodynamics with Antisymmetric Spin Tensors and an Extension of the Bargmann-Michel-Telegdi Equation}",
    eprint = "2506.20698",
    archivePrefix = "arXiv",
    primaryClass = "nucl-th",
    doi = "10.1103/59j2-5sm2",
    journal = "Phys. Rev. Lett.",
    volume = "136",
    number = "18",
    pages = "182301",
    year = "2026"
}

@article{Florkowski:2024cif,
    author = "Florkowski, Wojciech",
    title = "{Spin hydrodynamics}",
    eprint = "2411.19673",
    archivePrefix = "arXiv",
    primaryClass = "hep-ph",
    doi = "10.1016/j.jspc.2025.100028",
    journal = "J. Subatomic Part. Cosmol.",
    volume = "3",
    pages = "100028",
    year = "2025"
}

@article{Huang:2024ffg,
    author = "Huang, Xu-Guang",
    title = "{An introduction to relativistic spin hydrodynamics}",
    eprint = "2411.11753",
    archivePrefix = "arXiv",
    primaryClass = "nucl-th",
    doi = "10.1007/s41365-025-01784-3",
    journal = "Nucl. Sci. Tech.",
    volume = "36",
    number = "11",
    pages = "208",
    year = "2025"
}

@article{Bhadury:2024ckc,
    author = "Bhadury, Samapan",
    title = "{Relativistic spin hydrodynamics with momentum- and spin-dependent relaxation time}",
    eprint = "2408.14462",
    archivePrefix = "arXiv",
    primaryClass = "hep-ph",
    doi = "10.1103/PhysRevC.111.034909",
    journal = "Phys. Rev. C",
    volume = "111",
    number = "3",
    pages = "034909",
    year = "2025"
}

@article{Florkowski:2018fap,
    author = "Florkowski, Wojciech and Kumar, Avdhesh and Ryblewski, Radoslaw",
    title = "{Relativistic hydrodynamics for spin-polarized fluids}",
    eprint = "1811.04409",
    archivePrefix = "arXiv",
    primaryClass = "nucl-th",
    doi = "10.1016/j.ppnp.2019.07.001",
    journal = "Prog. Part. Nucl. Phys.",
    volume = "108",
    pages = "103709",
    year = "2019"
}

@article{Kapusta:2019sad,
    author = "Kapusta, Joseph I. and Rrapaj, Ermal and Rudaz, Serge",
    title = "{Relaxation Time for Strange Quark Spin in Rotating Quark-Gluon Plasma}",
    eprint = "1907.10750",
    archivePrefix = "arXiv",
    primaryClass = "nucl-th",
    doi = "10.1103/PhysRevC.101.024907",
    journal = "Phys. Rev. C",
    volume = "101",
    number = "2",
    pages = "024907",
    year = "2020"
}

@article{Ayala:2019iin,
    author = "Ayala, Alejandro and De La Cruz, David and Hern{\'a}ndez-Ort{\'\i}z, S. and Hern{\'a}ndez, L. A. and Salinas, Jordi",
    title = "{Relaxation time for quark spin and thermal vorticity alignment in heavy-ion collisions}",
    eprint = "1909.00274",
    archivePrefix = "arXiv",
    primaryClass = "hep-ph",
    doi = "10.1016/j.physletb.2019.135169",
    journal = "Phys. Lett. B",
    volume = "801",
    pages = "135169",
    year = "2020"
}

@article{Ayala:2023vgv,
    author = "Ayala, Alejandro and Bernal-Langarica, Santiago and Dom{\'\i}nguez Jim{\'e}nez, Isabel and Maldonado, Ivonne and Medina-Serna, Jos{\'e} Jorge and Rend{\'o}n, Javier and Tejeda-Yeomans, Mar{\'\i}a Elena",
    title = "{Relaxation time for the alignment between quark spin and angular velocity in a rotating QCD medium}",
    eprint = "2311.07859",
    archivePrefix = "arXiv",
    primaryClass = "hep-ph",
    doi = "10.1103/PhysRevD.109.074018",
    journal = "Phys. Rev. D",
    volume = "109",
    number = "7",
    pages = "074018",
    year = "2024"
}

@article{Ritus:1972ky,
    author = "Ritus, V. I.",
    title = "{Radiative corrections in quantum electrodynamics with intense field and their analytical properties}",
    doi = "10.1016/0003-4916(72)90191-1",
    journal = "Annals Phys.",
    volume = "69",
    pages = "555--582",
    year = "1972"
}

@unpiblished{Hernandez:2026lyd,
    author = "Hern{\'a}ndez, Luis A. and Zamora, R.",
    title = "{Scalar propagator in a rotating thermal medium: Consistency checks and applications}",
    eprint = "2609.04619",
    archivePrefix = "arXiv",
    primaryClass = "hep-ph",
    month = "9",
    year = "2026"
}

@article{Ambrus:2015lfr,
    author = "Ambrus, Victor E. and Winstanley, Elizabeth",
    title = "{Rotating fermions inside a cylindrical boundary}",
    eprint = "1512.05239",
    archivePrefix = "arXiv",
    primaryClass = "hep-th",
    doi = "10.1103/PhysRevD.93.104014",
    journal = "Phys. Rev. D",
    volume = "93",
    number = "10",
    pages = "104014",
    year = "2016"
}

\end{document}